%% file: main.tex
\documentclass[sigconf,nonacm]{acmart}

\usepackage{natbib}
\usepackage{multirow}
\usepackage{threeparttable}
\usepackage{arydshln}
\usepackage{pdflscape}
\usepackage{xstring}
\usepackage{listings}
\graphicspath{{./figures/}}

\begin{document}

\newcommand{\toolName}{XChainDataGen}

\newcommand{\rb}[1]{\textit{\color{ACMOrange}[RB] : #1}}
\newcommand{\andre}[1]{\textit{\color{ACMBlue}[AA] : #1}}
\newcommand{\mpc}[1]{\textit{\color{ACMRed}[MPC] : #1}}

\newcommand{\truncate}[1]{%
  \StrLeft{#1}{8}...\StrRight{#1}{8}
}

\newcommand{\AddrHref}[3][blue]{\href{#2}{\color{#1}{\truncate{#3}}}}%

\newcommand{\AddrHrefFootnote}[3][blue]{\href{#2}{\color{#1}{#3}}}%

\newcommand{\AddrHrefEthereum}[2][blue]{\href{https://etherscan.io/address/#2}{\color{#1}{\truncate{#2}}}}%
\newcommand{\AddrHrefArbitrum}[2][blue]{\href{https://arbiscan.io/address/#2}{\color{#1}{\truncate{#2}}}}%
\newcommand{\AddrHrefPolygon}[2][blue]{\href{https://polygonscan.com/address/#2}{\color{#1}{\truncate{#2}}}}%
\newcommand{\AddrHrefOptimism}[2][blue]{\href{https://optimistic.etherscan.io/address/#2}{\color{#1}{\truncate{#2}}}}%
\newcommand{\AddrHrefBase}[2][blue]{\href{https://basescan.org/address/#2}{\color{#1}{\truncate{#2}}}}%
\newcommand{\AddrHrefScroll}[2][blue]{\href{https://scrollscan.com/address/#2}{\color{#1}{\truncate{#2}}}}%
\newcommand{\AddrHrefLinea}[2][blue]{\href{https://lineascan.build/address/#2}{\color{#1}{\truncate{#2}}}}%
\newcommand{\AddrHrefRonin}[2][blue]{\href{https://app.roninchain.com/address/#2}{\color{#1}{\truncate{#2}}}}%
\newcommand{\AddrHrefGnosis}[2][blue]{\href{https://gnosisscan.io/address/#2}{\color{#1}{\truncate{#2}}}}%
\newcommand{\AddrHrefAvalanche}[2][blue]{\href{https://snowtrace.io/address/#2}{\color{#1}{\truncate{#2}}}}%

\newcommand{\Shorten}[1]{\truncate{#1}}%

\title{\toolName: A Cross-Chain Dataset Generation Framework}


\author{André Augusto}
\affiliation{%
  \institution{INESC-ID and IST, University of Lisbon}
  \country{Portugal}}

\author{André Vasconcelos}
\affiliation{%
  \institution{INESC-ID and IST, University of Lisbon}
  \country{Portugal}}

\author{Miguel Correia}
\affiliation{%
  \institution{INESC-ID and IST, University of Lisbon}
  \country{Portugal}}

\author{Luyao Zhang}
\affiliation{%
  \institution{Duke Kunshan University}
  \country{China}}

\renewcommand{\shortauthors}{Augusto A., et al.}

\begin{abstract}

The number of blockchain interoperability protocols for transferring data and assets between blockchains has grown significantly. However, 
no open dataset of cross-chain transactions exists to study interoperability protocols in operation. There is also no tool to generate such datasets and make them available to the community.

This paper proposes \toolName, a tool to extract cross-chain data from blockchains and generate datasets of cross-chain transactions (cctxs). Using \toolName, we extracted over 35 GB of data from five cross-chain protocols deployed on 11 blockchains in the last seven months of 2024, identifying 11,285,753 cctxs that moved over 28 billion USD in cross-chain token transfers.
Using the data collected, we compare protocols and provide insights into their security, cost, and performance trade-offs. As examples, we highlight differences between protocols that require full finality on the source blockchain and those that only demand soft finality (\textit{security}). 
We compare user costs, fee models,
and the impact of variables such as the Ethereum gas price on protocol fees (\textit{cost}).
Finally, we produce the first analysis of the implications of EIP-7683 for cross-chain intents, which are increasingly popular and greatly improve the speed with which cctxs are processed (\textit{performance}), thereby enhancing the user experience.
The availability of \toolName~and this dataset allows various analyses, including trends in cross-chain activity, security assessments of interoperability protocols, and financial research on decentralized finance (DeFi) protocols.

\end{abstract}



\maketitle

\section{Introduction}

Interoperability is increasingly used in the blockchain ecosystem, enabling the communication between multiple blockchains, i.e., \emph{cross-chain communication}. 
Blockchain interoperability allows for
reducing liquidity fragmentation~\cite{lehar2023liquidity}, thus fostering a more efficient decentralized financial ecosystem. Reputable organizations in the blockchain ecosystem -- such as Uniswap, Chainlink, and Coinbase -- have recognized interoperability as a critical aspect for the wide adoption of blockchain technology~\cite{chainlink_interop, uniswap_across_partnership,coinbase_interop}.

The blockchain ecosystem has evolved to include various networks (blockchains) that optimize different aspects of the blockchain trilemma~\cite{blockchain_trilemma}: scalability, decentralization, and/or security. This variety has resulted in an increasing demand for robust interoperability solutions, enabling users to transfer assets and data seamlessly. The most common mechanism that allows communication between two blockchains is a \emph{cross-chain token bridge}, or just \emph{bridge}. However, interoperability goes beyond the need to transfer assets. Interoperability plays a crucial role in addressing the scalability challenges of the so-called Layer 1 (L1) blockchains -- such as Ethereum -- which are notorious for their performance limitations (around 10-15 transactions per second) and high costs -- with transaction fees occasionally soaring to thousands/millions of USD\footnote{e.g., \AddrHrefFootnote{https://etherscan.io/tx/0x2c9931793876db33b1a9aad123ad4921dfb9cd5e59dbb78ce78f277759587115}{0x2c9931793876db33b1a9aad123ad4921dfb9cd5e59dbb78ce78f277759587115}}. To address these issues, Layer 2 (L2) solutions have been developed, where users can 
use blockchains with lower fees and faster processing times without altering the underlying trust assumptions of the L1 blockchains~\cite{l2_fees_info, gudgeon_sok_2020} -- i.e., they inherit L1's security. Communication between L1s and L2s is also facilitated by a \emph{bridge}.

\begin{figure}[tbh]
    \centering
    \includegraphics[width=\columnwidth]{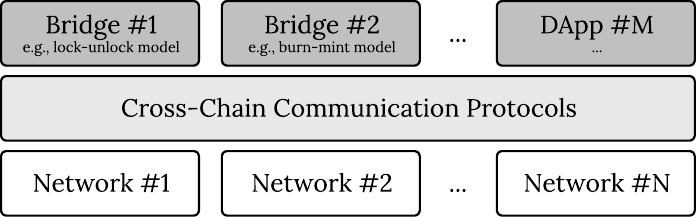}
    \caption{Cross-chain bridges built on top of cross-chain communication protocols.}
    \label{fig: amb-architecture}
    \Description{<long description>}
\end{figure}

In the past, the main problem with existing bridges was that they were designed in an ad hoc manner to solve specific interoperability issues, i.e., to interoperate specific pairs of blockchains. Then, Arbitrary Message Passing (AMP) protocols appeared as a class of generic messaging protocols that connect decentralized applications deployed on different blockchains, abstracting the complexity of cross-chain communication (cf. Figure~\ref{fig: amb-architecture}). 
These protocols allow the transfer of arbitrary data between blockchains\footnote{An example ``Hello World!'' message 
sent across chains can be found in~\cite{hello_world_example}.}, 
which may be forwarded to and interpreted by decentralized applications, such as cross-chain bridges. AMP protocols enable not only asset transfers but also the execution of arbitrary actions across a wide range of blockchains (e.g., cross-chain function calling).

With all these developments, there are some theoretical comparisons between protocols in the literature \cite{augusto_sok_2024, belchior2021survey}. However, there are no protocol comparisons based on \textit{empirical data}. Moreover, there is no open-source data that allows the research community to conduct such studies. We believe that there are no open datasets of cross-chain transactions due to the cumbersome process of extracting cross-chain data. \textit{This process is difficult due to many technical challenges,} starting with the extraction of data from multiple blockchains, to linking transactions and events emitted across them according to each protocol's custom logic. Not only do different protocols have completely different implementations, but bridges can follow different transferring models (lock-unlock, burn-mint, lock-mint) ~\cite{belchior_2023_doyouneed}, and distinct tokens are usually handled differently from bridge to bridge -- e.g., transferring an ERC-20 token (or any smart contract-based token) is different from transferring the native currency of a blockchain.

In recent years, cross-chain intents, now being standardized in the Ethereum Improvement Proposal (EIP) 7683~\cite{eip_7683}, have emerged as a promising solution to improve the user experience when using bridges. These mechanisms guarantee that \emph{cctxs} are completed in a matter of seconds -- a concrete example is illustrated in Section~\ref{subsec: cc_intents}. However, due to the novelty of this \emph{intent} paradigm, there is also scarce academic research in this area. We also fill this gap by providing the first empirical analysis of a cross-chain protocol based on cross-chain intents -- Across protocol~\cite{across}.

\label{subsec: rqs}

This paper has three main contributions. Each contribution answers three research questions:
\begin{itemize}

    \item \textit{RQ1: How can we design a scalable and reusable tool for extracting cross-chain data and generating cross-chain transactions?}  
    We introduce \textbf{\toolName}, a framework for generating cross-chain transaction datasets. Currently, our framework supports 8 bridges deployed across 11 blockchains, enabling large-scale empirical analysis of interoperability protocols. Using \toolName, we compiled \textbf{the first large-scale open dataset} of \textbf{11,285,753} \emph{cctxs} from 5 cross-chain bridges across 11 blockchains, built on AMP protocols during the last seven months of 2024. With \toolName, we obtain improved \textbf{reproducibility}. Section~\ref{sec: solution} details the framework. 

    \item \textit{RQ2: How do cross-chain bridges compare in terms of performance, cost, and security trade-offs?}  
    We present the \textbf{first comparative study of major cross-chain bridges}, examining their performance in processing transactions, the financial costs incurred by users, and their \textbf{security, cost, and performance trade-offs}. 
    Sections~\ref{sec: results} and \ref{sec: discussion} provide the analysis of a dataset generated using \toolName~and discuss key findings.

    \item \textit{RQ3: What are the key future research directions for cross-chain protocols with this new dataset and tool?}
    Building on our empirical findings and existing theoretical analyses, we outline \textbf{a set of critical research directions} for researchers, protocol developers, and users. These insights aim to guide advances in security and usability within interoperability protocols. Section~\ref{sec: discussion} explores these future research avenues.

\end{itemize}

All monetary values in this paper are in USD and estimated using the average daily token prices (sourced from Alchemy~\cite{alchemy}).

\section{Background}
\label{sec: background}

This section overviews some core concepts: cross-chain bridges, cross-chain transactions, and cross-chain intents.

\subsection{Cross-Chain Bridges \& Cross-Chain Txs}

Cross-chain bridges are decentralized applications that, contrary to other DApps, are deployed on two or more blockchains and facilitate the transfer of assets across them. They require off-chain mechanisms (sometimes referred to as \emph{bridge operator} or \emph{validator}) that either synchronize the state between smart contracts on multiple blockchains (e.g., using light-clients~\cite{Zamyatin_xclaim_2019}), or generate digital signatures later verified on-chain. The specific architectures and proving systems used vary from bridge to bridge~\cite{augusto_sok_2024}.

\begin{figure}[ht]
    \centering
    \includegraphics[width=0.48\textwidth]{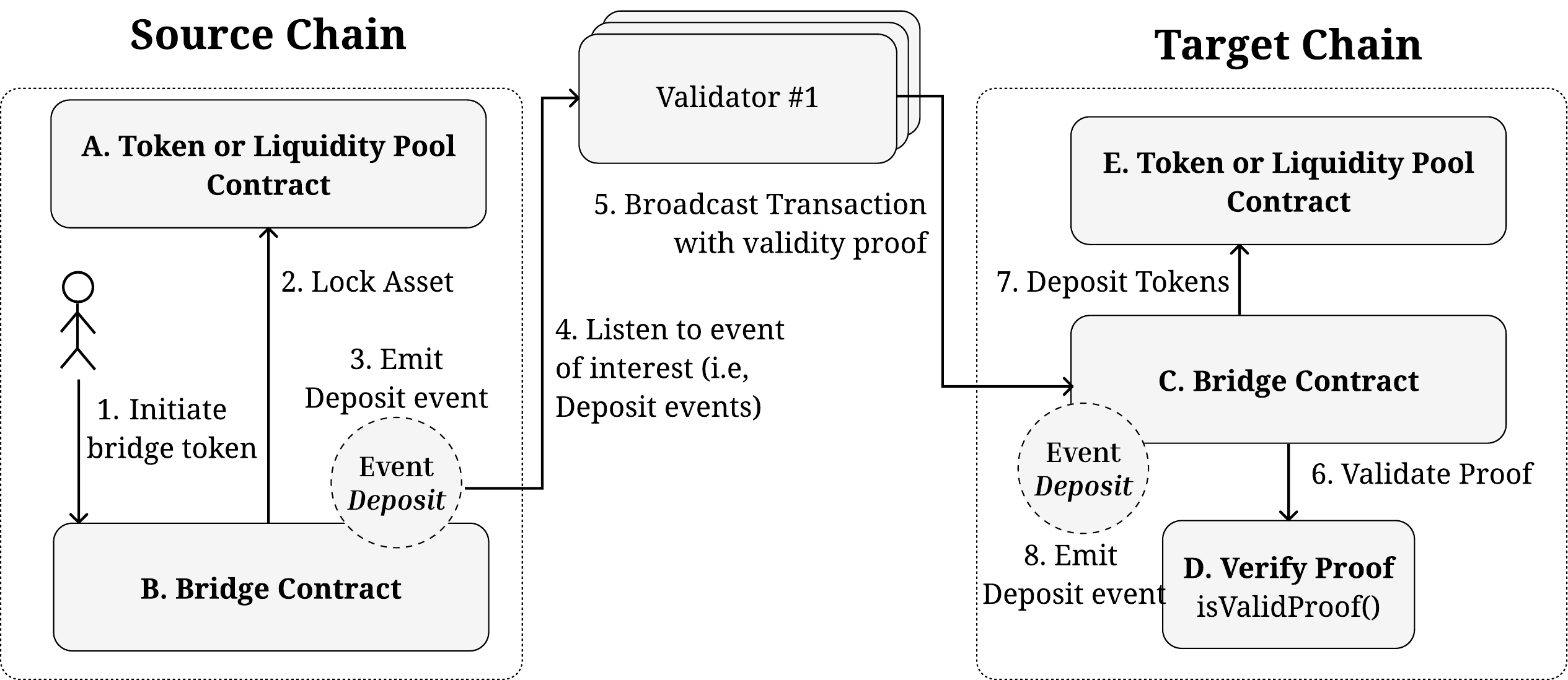}
    \caption{An example of the flow of information for a cross-chain token transfer using a \textit{bridge}.
    }
    \label{fig: bridge-flow}
    \Description{<long description>}
\end{figure}

Let \( B \) be the set of all blockchains supported by a cross-chain protocol, with \( |B| \geq 2 \), and let \( T_B \) represent the set of all transactions occurring on blockchains in \( B \). An example of the usual flow of information on a token bridge between blockchains $B_1$ and $B_2$ is in Figure~\ref{fig: bridge-flow}. Consider a user address (\textit{sender}) that interacts with the bridge contract on the source blockchain to deposit a certain amount (quantity $q_{a_{B_1}}$) of an asset $a_{B_1}$. This contract enforces the lock or burn of those assets owned by the user and emits an event signaling the success of this process in $B_1$. Off-chain entities listen to these events of interest and relay data and/or proofs to $B_2$, attesting to the validity of the deposit on $B_1$. If the proof is valid, the bridge contract triggers the deposit of a quantity $q_{a_{B_2}}$ of an asset $a_{B_2}$ to the recipient's address \emph{r}.

In this paper, we define a cross-chain transaction ($cctx$) as the union of transaction and event data from two blockchains. The set of all \emph{cctxs} between blockchains in \( B \), is defined as:

\[
\text{CCTXs} = \bigcup_{
    \substack{\text{tx}_{B_1} \in T_{B_1},\\ \text{tx}_{B_2} \in T_{B_2},\\ B_1 \neq B_2\\ }
} cctx(\text{tx}_{B_1}, \text{tx}_{B_2}, s, r, \text{id}, a_{B_1}, a_{B_2}, q_{a_{B_1}}, q_{a_{B_2}})
\]

where:
\begin{itemize}
    \item \( s \) and \( r \) are the sender and recipient addresses on \( B_1 \) and \( B_2 \), respectively;
    \item \( a_{B_1} \) and \( a_{B_2} \) are smart contracts representing assets on \( B_1 \) and \( B_2 \), respectively;
    \item \( q_{a_{B_1}} \) is the amount of \( a_{B_1} \) sent on \( B_1 \);
    \item \( q_{a_{B_2}} \) is the amount of \( a_{B_2} \) received on \( B_2 \);
\end{itemize}

A \textit{cross-chain swap} occurs when \( a_{B_1} \) and \( a_{B_2} \) represent different tokens, e.g., to deposit WETH on Ethereum and receive USDC on Arbitrum. In the case of \textit{batching}, the same transaction \( \text{tx}_{B_i} \) on the blockchain \( B_i \) may be associated with multiple \emph{cctxs}, if it triggered multiple token transfers, on either blockchain. In this case, the cost of issuing a blockchain transaction is amortized by the number of \emph{cctxs} it is involved in.

\subsection{Cross-Chain Intents and EIP 7683}
\label{subsec: cc_intents}

\begin{figure}[ht]
    \centering
    \includegraphics[width=0.48\textwidth]{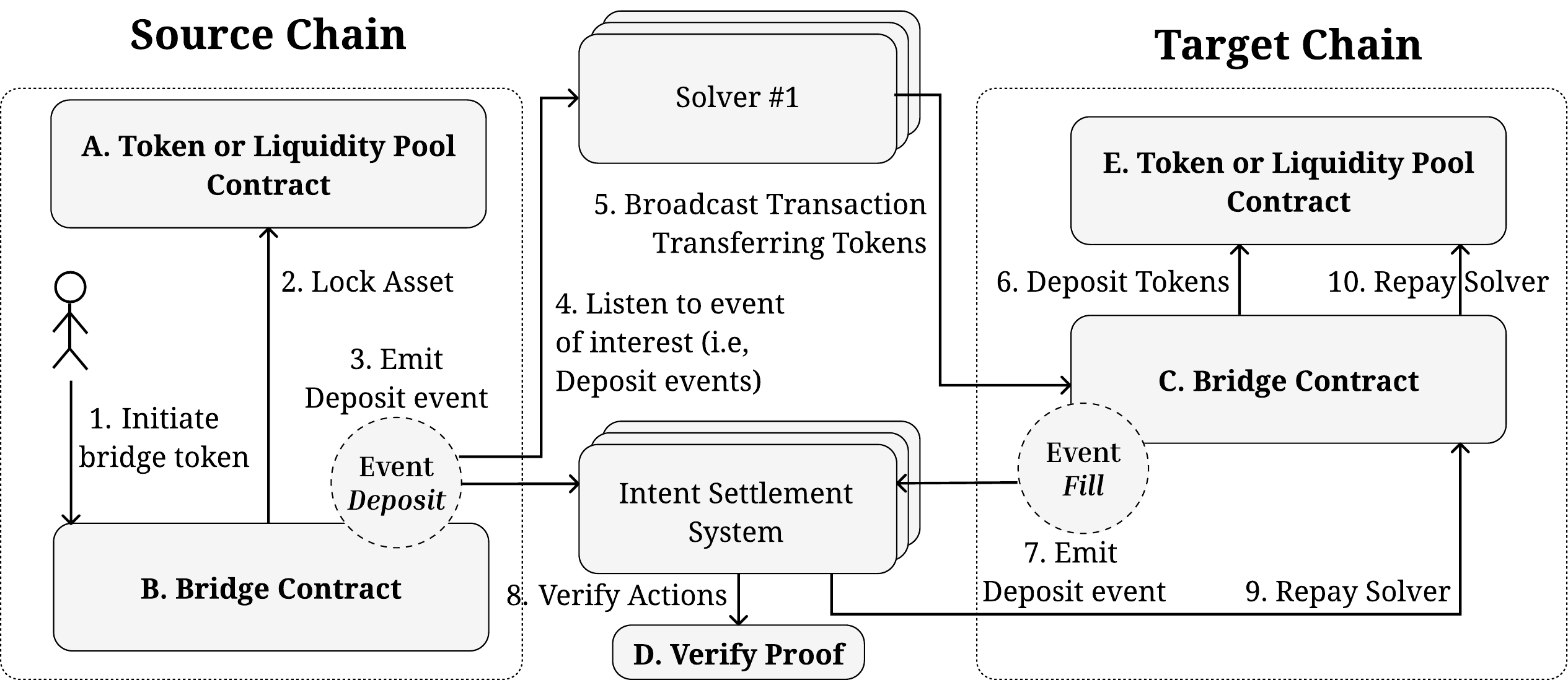}
    \caption{Example information flow for a cross-chain token transfer using a \textit{bridge} powered by cross-chain intents.
    }
    \label{fig: bridge-flow-intents}
    \Description[<short description>]{<long description>}
\end{figure}

Cross-chain intents, now standardized in EIP-7683~\cite{eip_7683}, allow users to express their desire (intent) to perform a cross-chain action without directly managing the transaction details across multiple blockchains. Once an intent is submitted, a network of entities called \textit{solvers} competes in a Dutch auction-style process~\cite{chitra2024analysisintentbasedmarkets} by paying upfront costs to fill the user's order on the destination blockchain \textbf{immediately}. Later, an off-chain \textit{intent settlement system} gathers events from both blockchains to verify that the solver correctly filled the user's order and issues repayments to the solver with a profit margin -- usually, the repayment can be performed on any blockchain according to the solver's preference. The Across interoperability protocol is the most advanced implementation of this paradigm; Across recently partnered with Uniswap for cross-chain swaps~\cite{uniswap_across_partnership}. Figure~\ref{fig: bridge-flow-intents} depicts the workflow and how it differs from the usual flow of a token bridge (cf. Figure~\ref{fig: bridge-flow}).

\section{\toolName}
\label{sec: solution}

\begin{figure*}
    \centering
    \includegraphics[width=0.95\textwidth]{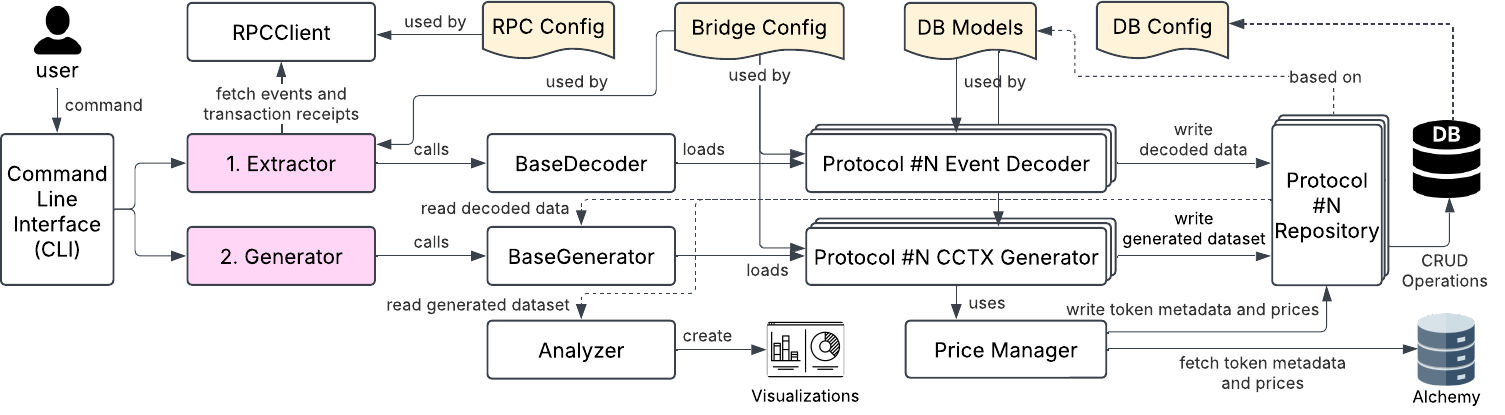}
    \caption{Architecture of \toolName~-- an open source framework for building cross-chain datasets. The framework can be extended to support additional bridges, requiring the protocol configuration file (which specifies the relevant contracts and events to capture in each contract for each blockchain) and the implementation of a custom decoder, generator, and repository.}
    \label{fig: architecture}
    \Description[<short description>]{<long description>}
\end{figure*}

\toolName\footnote{a simplified and anonymized repository can be found in \href{https://anonymous.4open.science/r/XChainDataGen-CC5B}{https://anonymous.4open.science/r/XChainDataGen-CC5B}}, represented in Figure~\ref{fig: architecture}, is a modular and extensible framework for the \textbf{extraction} of cross-chain data and \textbf{generation} of cross-chain transactions. Currently, \toolName~supports 9 cross-chain bridges: CCTP by \textit{Circle}~\cite{cctp}, CCIP by \textit{Chainlink}~\cite{ccip}, Stargate Finance (4 modes: taxi mode~\cite{stargate_taxi_and_bus}, bus mode~\cite{stargate_taxi_and_bus}, hydra mode~\cite{stargate_hydra}, and cross-chain swaps~\cite{stargate_swaps}), Omnibridge~\cite{omnibridge}, XDAI bridge~\cite{xdai_bridge}, Polygon PoS~\cite{polygon_portal}, Ronin Bridge~\cite{ronin_bridge}, and Across~\cite{across}.

\subsection{Command Line Interface (CLI)}

\toolName~provides a command line interface (see Listing~\ref{listing:cli}) with two possible actions, extract (\S\ref{subsec: extractor}) and generate (\S\ref{subsec: generator}). \toolName~first retrieves all events from the bridge contracts across the specified blockchains. Second, it loads the extracted data in the last step and generates \emph{cctxs} by linking data across blockchains. A public Docker image of the tool will be made available.

\begin{lstlisting}[language=bash, caption={Command Line Interface for \toolName}, label=listing:cli]
XChainDataGen.py {extract, generate}
     --bridge <BRIDGE>
     --start_ts <START_UNIX_TS> --end_ts <END_UNIX_TS>
     --blockchains <CHAIN 1> <CHAIN 2> ... <CHAIN N>
\end{lstlisting}

\subsection{Artifacts}

\toolName~is initialized with a set of configuration files:
\begin{itemize}
    \item \textbf{Database Configs}: Define the connection parameters for the data storage to be used.
    \item \textbf{Database Models}: Each bridge has its own events and event models. For each bridge, we retrieved event models from the contracts' Application Binary Interface (ABI).
    \item \textbf{Bridge Configs}: Defines the mapping between smart contracts and the events to be captured for each contract on each supported blockchain. These configuration files are based on the official bridge documentation, examining deployed contracts, and interacting with the protocols\footnote{e.g., the config file for Stargate is in \href{https://anonymous.4open.science/r/XChainDataGen-CC5B/extractor/stargate/constants.py}{https://anonymous.4open.science/r/XChainDataGen-CC5B/extractor/stargate/constants.py}}.
    \item \textbf{RPC Configs}: Contains a list of RPC URLs for the supported blockchains. These are used 
    in a round-robin fashion to fetch data from blockchain nodes.
\end{itemize}

\subsection{Extractor}
\label{subsec: extractor}

The \textit{Extractor} takes as input (i) the bridge to be analyzed, (ii) a time interval defined using Unix timestamps, and (iii) a set of supported blockchains. The extraction process works as follows. First, it loads the bridge configuration file, which specifies all relevant contract events for each blockchain where the bridge is deployed. 
The Extractor iterates over the user-specified blockchains, determining the nearest block numbers corresponding to the provided timestamps (i.e., start and end blocks for each blockchain). Then, it divides this block range into intervals of 2,000 blocks\footnote{usually, the maximum supported by public RPC nodes} and retrieves logs for all specified events in each contract using the \texttt{eth\_getLogs} RPC method. For every captured event, the Extractor also fetches the corresponding transaction receipt and block information using \texttt{eth\_getTransactionReceipt} and \texttt{eth\_getBlockByNumber}. Each event is decoded using a base decoder or, when necessary, a custom decoder tailored to the specific contract and event type. The extracted data is stored in a storage system, with each event written as a separate relation. We implemented a Repository Pattern -- abstracting the data layer and allowing different storage systems -- to guarantee flexibility. Users can customize the storage system based on their specific dataset requirements through the database configuration file.
At the end of the extraction phase, the storage system contains all the data associated with the specified bridge events and blockchains.

\subsection{Generator}
\label{subsec: generator}

The goal of the \textit{Generator} is to build cross-chain transactions based on previously extracted data. One of the main challenges of analyzing cross-chain protocols is their ad-hoc architecture. Therefore, because of the specific logic and interfaces of each one, it was hard to create a base generator with the majority of the logic. To cope with this difficulty, the base generator dynamically loads a custom generator for the bridge to be analyzed. The custom generator reads the data previously extracted and written to the storage system and merges the different records to create cross-chain transactions. Records are merged based on cross-chain transaction identifiers (deposit and withdrawal IDs, or message IDs according to the protocol). Beyond these identifiers, we insert checks based on the sender, recipient, token addresses, and amounts transferred. The specific fields through which records are merged depend on the logic of each bridge and the data model of each event. At the end of this phase, the storage system also contains datasets of cross-chain transactions.

\section{Empirical Analysis of Cross-Chain Data}
\label{sec: results}

We used \toolName~to extract data from 5 cross-chain bridges deployed in 11 different blockchains (not all bridges supporting all blockchains) -- Ethereum, Avalanche, Binance Smart Chain, Arbitrum, Base, Optimism, Linea, Scroll, Polygon, Ronin, and Gnosis -- in the last 7 months of 2024 (Jun 1, 2024 to Dec 31, 2024). Using this data, we generated a data set of over \textbf{11 million cross-chain transactions} that moved a total value of \textbf{28.20 billion USD}. Table~\ref{table: summary} summarizes the extracted data and provides an overview of the protocol fees charged to the users. In addition to the fees paid to the protocol, the user pays the transaction fee for the initial transaction on the source blockchain, and the bridge operator pays the transaction fee for the final transaction on the destination blockchain (see Figure~\ref{fig: bridge-flow}).

\subsection{Bridges Analyzed}

This section reviews the cross-chain bridges analyzed in our study. Due to size constraints, we do not show data related to all protocols currently supported by \toolName. The selection of these projects was driven by 1) their Total Value Locked (TVL), based on data from L2Beat and DefiLlama~\cite{l2beat, defillama}, and 2) the need to capture a diverse set of underlying assumptions -- at least a centralized bridge, one employing a batching mode, and a cross-chain intent-based solution.

\input{tables/summary.tex}

\subsubsection{Cross-Chain Transfer Protocol (CCTP) by Circle}

CCTP~\cite{cctp} was specifically designed to facilitate native transfers of USDC across blockchains. According to the project documentation, before signing any attestation for a source chain event, the off-chain component waits for the finality of the deposit transaction. 
Circle's centralized attestation service issues the transaction on the destination blockchain, making the protocol inherently centralized.

\subsubsection{Cross-Chain Interoperability Protocol (CCIP) by Chainlink}

Chainlink's CCIP~\cite{ccip} employs a multi-layered architecture to ensure secure cross-chain communication. The protocol has two primary off-chain networks: 1) a committing network that relays data between blockchains; and 2) an execution network that verifies and attests to the relayed data. In addition, a dedicated risk management network oversees operations on both chains, ensuring that actions comply with the expected behavior of the system. Although separating concerns among these components increases security, it contributes to higher latency in transaction processing.

\subsubsection{Stargate Finance (Taxi Mode)}

Stargate Finance~\cite{stargate_taxi_and_bus}, the first decentralized application built on Layer Zero, utilizes liquidity pools on both sides of the bridge to enable seamless cross-chain interactions. According to the documentation, in Taxi mode, the protocol processes one-to-one transactions immediately once the source chain reaches finality (we later show that the protocol only waits for \emph{soft} finality~\cite{10.1145/3631310.3633493}).

\subsubsection{Stargate Finance (Bus Mode)}

Stargate Finance~\cite{stargate_taxi_and_bus} introduced the Bus Mode in V2, a cross-chain transaction batching mechanism designed to reduce costs. In this mode, users initiate transactions on the source blockchain, and once a timeout is reached or a fixed number of transactions are gathered (i.e., all the bus seats are filled), the protocol dispatches the ``bus''. This process aggregates all token transfers into a single consolidated transaction on the source chain, assigns a unique identifier to the batch, and relays the bundled data to the destination blockchain. On the destination blockchain, one blockchain transaction triggers all necessary token transfers to the users' addresses.

\subsubsection{Across Protocol}

Across~\cite{across} is a cross-chain protocol that leverages cross-chain intents, in line with EIP-7683. Across allows very fast transfers of value across blockchains by fulfilling users' orders (i.e., requests to bridge funds) in seconds without waiting for finality on the source blockchain. We include Across in our analysis because this new paradigm has the potential to revolutionize the user experience during cross-chain interactions.

\subsection{Performance and Cost Empirical Analysis}

In our empirical study, we evaluated performance and cost metrics in 54 pairs of blockchains\footnote{complete tables in \href{https://anonymous.4open.science/r/xchaindatagen-analysis-4085/}{https://anonymous.4open.science/r/xchaindatagen-analysis-4085/}}. Given the vast amount of data collected, we categorized each blockchain as either Layer 1 (L1) or Layer 2 (L2)\footnote{Layer 2 blockchains include a wide range of solutions, including payment channels, state channels, and commit-chains~\cite{gudgeon_sok_2020}. In this work, when referring to L2 blockchains, we specifically focus on commit-chains -- namely rollups -- built on top of Ethereum.}, and structured our analysis around the transfer direction, considering the following cases: L1 $\rightarrow$ L1, L1 $\rightarrow$ L2, L2 $\rightarrow$ L1, and L2 $\rightarrow$ L2. For each transfer direction, we selected one or two representative blockchain pairs -- those with the highest volume of token transfers and supported by all protocols -- to serve as examples. Furthermore, due to the high dispersion of the data (later shown in Figure~\ref{fig: l2_l2_latency_distribution}), we focus our analysis on the 25th (Q1), 50th (Q2), and 75th percentile (Q3) of the distributions, also including the Interquartile Range (IQR).

\subsubsection{L1 $\rightarrow$ L1 Transfers}
\label{section: l1-l1}

Across is left out of this comparison because the only Layer 1 blockchain supported by the protocol is Ethereum, so there is no possible L1 $\rightarrow$ L1 pair. The only blockchain pair supported by all the other protocols is Ethereum to Avalanche. Table~\ref{table: l1-l1} summarizes the overall latency and cost results for token transfers between these blockchains, and we summarize the main results below.

\input{tables/L1-L1-latency-cost.tex}

We first focus on token transfers from Ethereum to Avalanche. Both CCIP and CCTP wait for the source blockchain to reach finality. Both medians are around 1000 seconds (16 minutes), and we captured minimum latencies of 809 seconds for CCIP and 787 seconds for CCTP, which align closely with Ethereum finality requirement of approximately two epochs (roughly 13 minutes, or about 780 seconds). There is a huge disparity between these values compared to the latency of Stargate Taxi mode, which is due to the enforced waiting period for source blockchain finality -- the first two wait for the \textit{full} finality of Ethereum, whereas Stargate Taxi mode employs a much shorter waiting period of around 200 seconds -- employing \textit{soft} finality, with fewer security guarantees. In the opposite direction, we cannot clearly identify this dependency on the source chain finality period. That is, the latency for transactions from Ethereum to Avalanche differs significantly from that in the reverse direction. This disparity is highly attributable to the longer finality period of Ethereum ($\approx$13 minutes) compared to Avalanche’s ($\approx$1 second). When transferring from Ethereum to Avalanche, Stargate Bus mode incurs roughly twice the latency of the Taxi mode; from Avalanche to Ethereum, the Bus mode latency increases by a factor of 4.5 -- i.e., unlike all other protocols, the batching mode is worse performant when dealing with a faster source blockchain.

Regarding the user costs, when transferring from Ethereum to Avalanche, there is a notable difference between CCTP and Stargate Taxi, with costs averaging \$5.56 and \$15.23, respectively. Neither protocol requires users to pay protocol fees (see Table~\ref{table: summary}), but transactions conducted through Stargate Taxi consistently incur significantly higher expenses, as detailed in subsequent subsections. We investigated the correlation between transaction timestamps and user costs to see if periods of elevated Ethereum gas prices were associated with increased Stargate activity, but our hypothesis was not confirmed. The findings indicate that throughout the entire analysis period, Stargate Taxi remained more expensive than CCTP, suggesting that the costs associated with interacting with the CCTP contract are lower than those for the Stargate contract. Additionally, transactions utilizing the Stargate Taxi mode often involve computations from third-party protocols (such as Decentralized Exchanges or Aggregators~\cite{Subramanian_2024}), which help explain the higher costs associated with Stargate transactions. The Stargate bus mode is slightly cheaper than CCTP, with a median cost of \$4.47. CCIP imposes an expensive flat fee with a median value of \$14.87.

\begin{figure}[ht]
    \centering
    \includegraphics[width=0.47\textwidth]{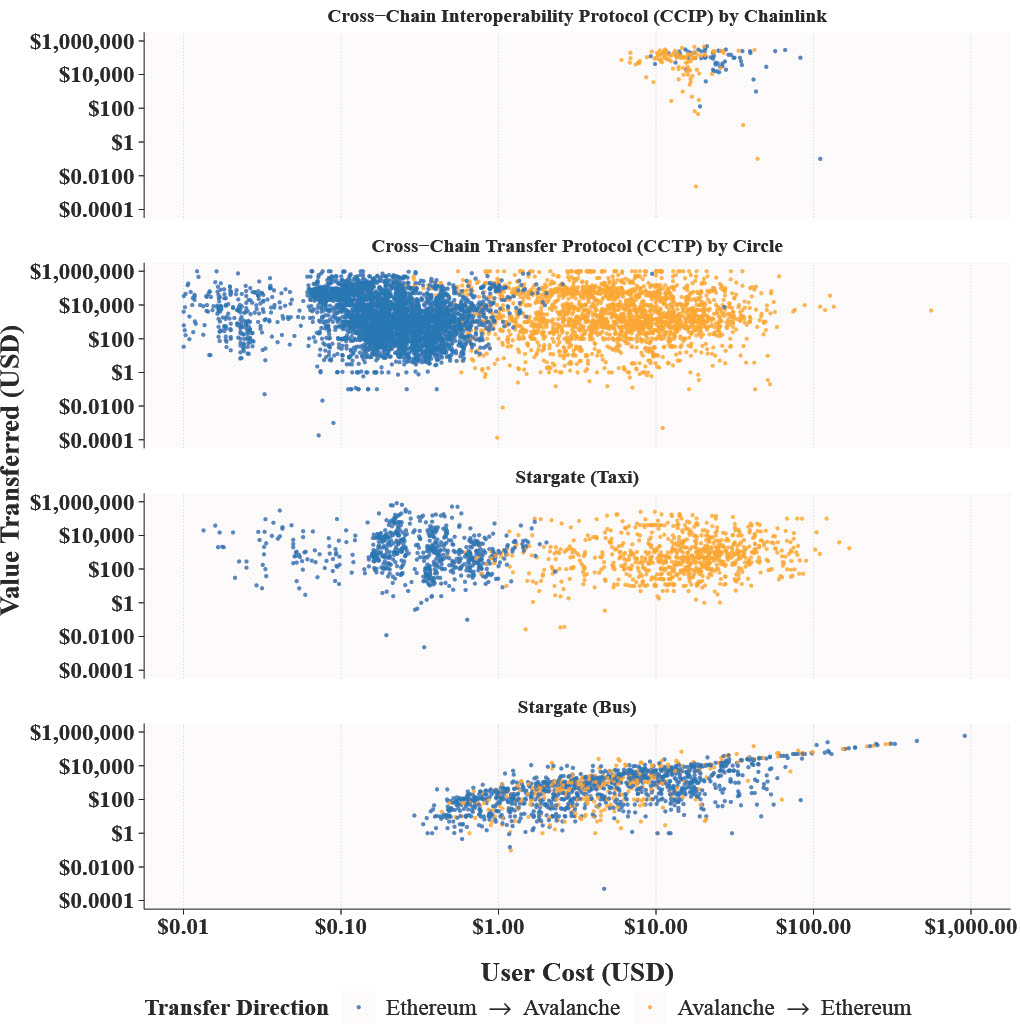}
    \caption{A comparison of the cost incurred by users versus the value transferred (in USD) for every cross-chain transaction between Ethereum and Avalanche.}
    \label{fig:comparison_eth_avax_cost}
    \Description{[Short description of the figure.]}
\end{figure}

When transferring from Avalanche, the cost to the users varies significantly. Given that issuing transactions on Avalanche is cheaper than on Ethereum (and users pay the source transaction fee), we would expect a reduction in overall costs. However, with CCIP, users actually encounter increased charges -- a substantial flat token fee. This is due to the users being charged the anticipated expenses associated with the operator transaction on Ethereum. Furthermore, Stargate Bus mode proves to be considerably more expensive compared to Stargate Taxi and CCTP. The main conclusion is that batching becomes slower and more expensive when transferring tokens from cheap and fast blockchains, as there are not that many costs to begin with. Further research showed another reason the Stargate Bus mode is more expensive than CCTP and Stargate's Taxi mode. Figure~\ref{fig:comparison_eth_avax_cost} shows the correlation between user costs and transfer value, with color coding that indicates the source blockchain. Unlike other protocols, the protocol fee charged to users in Stargate Bus mode depends on the value transferred. The data also confirms that it is generally more cost-effective for users to bridge from chains with lower transaction fees, such as Avalanche, rather than from those with higher fees. Stargate Bus mode not only demonstrates a clear dependence on the transferred value but also shows less sensitivity to the specific source and destination chains. 

\subsubsection{L1 $\rightarrow$ L2 Transfers}
\label{section: l1-l2}

In this subsection, we analyze two examples from L1 to L2 blockchains -- the highest in terms of volume transacted: Ethereum to Base and Avalanche to Base. Table~\ref{table: l1-l2} shows the latency and cost metrics for these transfers.

\input{tables/L1-L2-latency-cost}

Interestingly, the results are similar to those in Table~\ref{table: l1-l1} (L1 $\rightarrow$ L1). For the Ethereum to Base pair, the cross-chain latency for CCTP and CCIP is around 1,060 seconds (approximately 17.5 minutes). In contrast, Stargate operating in Taxi mode is the fastest solution, with 50\% of transactions completed in 210-224 seconds. Although Stargate's Bus mode requires roughly double the time -- largely due to the batching process -- just like in L1 $\rightarrow$ L1, it still outperforms both CCTP and CCIP, which wait for the first transaction's finality on Ethereum plus additional processing time. The Across protocol, which uses an intent-based model, exhibits the shortest latencies, with half of its cross-chain transactions completing between 16 and 28 seconds -- showcasing right away the potential of cross-chain intents.

When considering the Avalanche to Base pair, the dynamics decrease notably. Given that Avalanche achieves finality much faster than Ethereum, the overall latencies are considerably reduced. However, the speed of each bridge still depends on the underlying processing mechanisms. In this scenario, Stargate's Bus mode falls behind, displaying roughly four to five times worse performance when compared to the other three protocols -- all three present very similar latencies separated by 10 to 20 seconds. Stargate Bus's longer latencies in both pairs are counterbalanced by significantly lower user costs -- it records a median cost of \$2.56 in Ethereum $\rightarrow$ Base. Across and CCTP show median costs of \$3.47 and \$3.51, respectively. CCIP and Stargate Taxi, similarly to the L1 $\rightarrow$ L1 case, show a higher cost, following the same rationale explained in Section~\ref{section: l1-l1}.

\subsubsection{L2 $\rightarrow$ L1 Transfers}
\label{section: l2-l1}

Table~\ref{table: l2-l1} shows the latency and cost results for transferring from L2 $\rightarrow$ L1 using Base to Ethereum and Base to Avalanche as examples. Having an L2 blockchain as the source chain could signify that the latency would be much slower. However, recall that L2s are built on top of L1s and -- thereby depending on the L1. In the table, we do not see any improvements in the cross-chain latency in both CCTP and CCIP that wait for the full finality of the transactions on the source blockchain -- in fact, we see an increase of 300-400 seconds compared to L1 $\rightarrow$ L2, which is the time taken for the L2 state to be aggregated, sequenced, and published on the L1~\cite{gudgeon_sok_2020}. Half of the data points are between 1,200 and 1,500 seconds, depending on the blockchain pair.

On the other hand, Stargate's two modes and Across present lower latencies in this setting compared to when the source blockchain is an L1. However, similarly to when transferring from L1s, Stargate Bus offers a slower solution, followed by Stargate taxi mode, and finally Across, with 50\% of the data between 4 and 12 seconds. Among these three protocols, there is a slight difference (approximately 80 seconds) when transferring to Ethereum versus Avalanche, with transactions targeting Ethereum experiencing longer delays. We hypothesize that this disparity is related to the speed of these blockchains: Ethereum has a slower block rate, resulting in transactions taking longer to be selected into a block. This issue can be deepened if the bridge operator opts to issue the final transaction on the destination blockchain with a low fee (to reduce costs), potentially causing the transaction to fall behind in the priority queue of block builders for the next blocks~\cite{eskandari2020sok}.

\input{tables/L2-L1-latency-cost}

Across's speed is a trade-off chosen by the protocol. Across has the second-highest median fee of 6.71\$ when transferring to Ethereum. CCIP leads the leaderboard with an impressive median of 26\$, following the trend highlighted in the last subsections. When transferring from Base to Ethereum, the protocol is required to make a transaction on Ethereum, making the flat fee set by the protocol increase drastically, increasing their profit margin -- the same conclusion can be drawn from the other pair Base to Avalanche (2.38\$ -- the second-highest fee in this configuration is Stargate's bus mode with 0.28\$, which is 8.5 times lower). CCTP and Stargate taxi mode only require the user to pay the transaction fee on the source chain, which is very cheap, keeping the values to 0.1 cents or fewer for both blockchain pairs.

\subsubsection{L2 $\rightarrow$ L2 Transfers}
\label{section: l2-l2}

Due to the low fees and high throughput on L2 blockchains, cross-chain interactions are more prevalent between them (68.90\% of all \emph{cctxs} collected are between L2s). In our study, we focus on the fund flow between Arbitrum and Base -- identified as the pair with over 10\% of the total cross-chain transactions in our dataset. Table~\ref{table: l2-l2} summarizes the results.

\input{tables/L2-L2-latency-cost}

\begin{figure}[ht]
    \centering
    \includegraphics[width=0.49\textwidth]{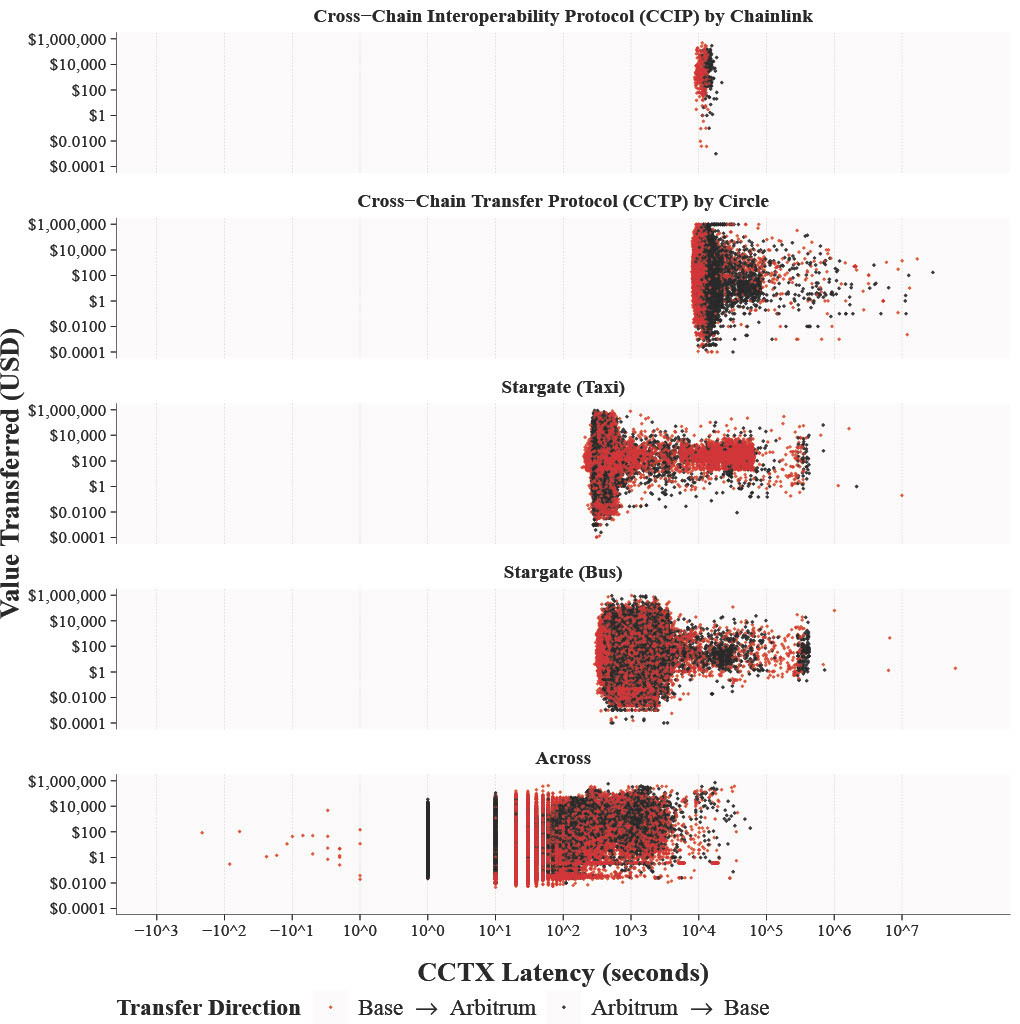} 
    \caption{Comparison between the latency versus the value transferred (in USD) per \emph{cctx} between Arbitrum and Base.}
    \label{fig: l2_l2_latency_distribution}
    \Description[<short description>]{<long description>}
\end{figure}

The latency data in CCTP and CCIP, when transferring from Base, is very similar to what was found in the last section and Table~\ref{table: l2-l1}, with the median latency around 1200-1500 seconds -- i.e., waiting for the full finality of the source blockchain. In the other protocols, we see a massive increase in performance, and a stabilization of all values in both blockchain pairs -- i.e., for Stargate's both modes and Across, the latency values are very similar regardless of the Layer 2 blockchains chosen as source and destination. Figure~\ref{fig: l2_l2_latency_distribution} backs this statement, where, contrary to Table~\ref{table: l1-l1}, there is no clear trend based on the source blockchain.

The performance of the Across protocol stands out. Our data indicate that 49.23\% of its cross-chain transactions are completed within 3 seconds, and 73.42\% within 10 seconds. Although most of the values are within these small periods, there are cases where transaction latencies extend significantly, with some cross-chain transactions recording latencies exceeding $10^3$ seconds and, in a couple of cases, even $10^4$. These results highlight the potential benefits of adopting cross-chain intents as a new standard for DeFi protocols in terms of performance and scalability when prioritizing user experience. The second most intriguing finding is the location of negative-latency cross-chain transactions, meaning that the transaction on the target blockchain was accepted first rather than the corresponding transaction on the source chain. As an example, consider Across's cross-chain transaction with id \emph{1996859} transferring $\approx$0.12 WETH from Arbitrum (\AddrHref{https://arbiscan.io/tx/0x8342a9ba82f6706c80faf7dd621d6cada11f479949b0756781edbfbd01d3cd61}{0x8342a9ba82f6706c80faf7dd621d6cada11f479949b0756781edbfbd01d3cd61}) to Base (\AddrHref{https://basescan.org/tx/0xdcb9f570dfd99a7b576e48dd67263613e38c81d9cd1aeca80562263f00fc037c}{0xdcb9f570dfd99a7b576e48dd67263613e38c81d9cd1aeca80562263f00fc037c}). The transaction on Base was accepted 7 seconds before the one on Arbitrum. We found various possible explanations for this behavior: 1) there is some privileged knowledge within the protocol probably through the interaction of the user with the User Interface; 2) \textit{solvers} are monitoring the public mempool looking for transactions that have not been added to a block yet, but will eventually; 3) there are \textit{solvers} with exclusivity for certain intents known before the first transaction is put in a block (following option 2).

The cost analysis also reveals interesting trends in the protocol fees. An unexpected finding in our analysis is the occurrence of negative costs in certain instances within the Stargate Bus protocol and in Across. These negative values indicate scenarios where users profit from executing cross-chain token swaps. In such cases, a user might input one unit of token $X$ with valuation $\phi_{\mathcal{B}_1}({X})$ in $\mathcal{B}_1$ and receive another token $Y$ in $\mathcal{B}_2$, at a relatively higher valuation -- i.e., $\phi_{\mathcal{B}_1}({Y})/\phi_{\mathcal{B}_1}({X}) < \phi_{\mathcal{B}_2}({Y})/\phi_{\mathcal{B}_2}({X})$. The computed ratio on the source blockchain is lower than on the destination blockchain, suggesting a potential arbitrage opportunity~\cite{qin_quantifying_2022, ferreira2024rolling}. We present a couple of examples: 1) a user inputs 5 USDC in Optimism (\AddrHref{https://optimistic.etherscan.io/tx/0x3bccadaf005cf1583258b0e7690e8527b9dd3c35fc98c6f0debd5535ed08dfa0}{0x3bccadaf005cf1583258b0e7690e8527b9dd3c35fc98c6f0debd5535ed08dfa0}) and the protocol outputs the equivalent of 5.45 USDC in another token on Base (\AddrHref{https://basescan.org/tx/0x1d01ffe42eeb452d71384e3bd858c1fbc405a5481a7517ad17742a7b11c554c3}{0x1d01ffe42eeb452d71384e3bd858c1fbc405a5481a7517ad17742a7b11c554c3}); 2) the user deposits 74 USDC on Optimism (\AddrHref{https://basescan.org/tx/0xdfc9c3463a8d9cba6a908be90a365058a56f56c23ee9d5783525250e28a723b3}{0xdfc9c3463a8d9cba6a908be90a365058a56f56c23ee9d5783525250e28a723b3}) and is able to withdraw around 2000 USD worth of tokens in Arbitrum (\AddrHref{https://basescan.org/tx/0xb21c43827243efcc6f5f392baed5bbc2d7ecf00f4bfa32d56c4ea4c4ecfed05f}{0xb21c43827243efcc6f5f392baed5bbc2d7ecf00f4bfa32d56c4ea4c4ecfed05f}). This phenomenon may be related to inaccuracies in token valuation estimates performed by price oracles on the source blockchain, which opens an interesting avenue for further research on price oracle manipulation~\cite{10646773} in this context. We found 1,285 transactions that made a profit for the user using Across, where 1,277 were cross-chain transactions from Optimism, suggesting a potential vulnerability in a contract related to the token price registry.

\begin{figure}[ht]
    \centering
    \includegraphics[width=0.49\textwidth]{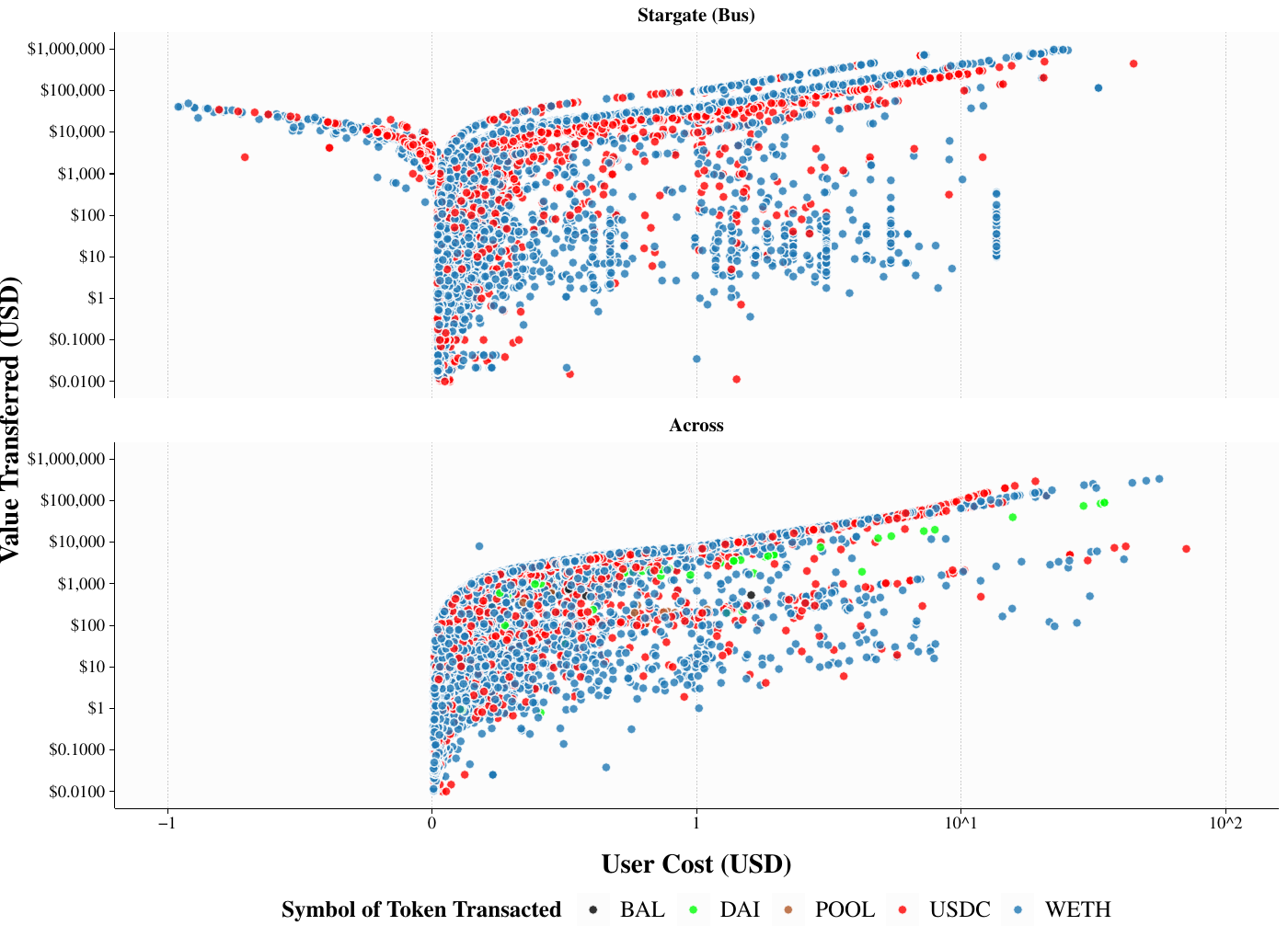}
    \caption{Correlation between User Cost and Value Transferred per \emph{cctx} in Stargate (Bus) and Across for \emph{cctxs} between Arbitrum and Base.}
    \label{fig: arb_base_cost_per_token}
    \Description[<short description>]{<long description>}
\end{figure}

We also present a plot, in Figure~\ref{fig: arb_base_cost_per_token}, with the correlation between user cost and transferred value. The only protocols that showed an interesting correlation were Stargate Bus and Across, and due to lack of space we opted to leave the others out. In the figure, it is noticeable multiple ascending linear trends (note that both axes are on a logarithmic scale), indicating that the cost is, at least, a function of the amount being transferred (and token represented by the color coding). With further analysis, we identified that different slopes correspond to different timeframes (e.g., in Stargate Bus mode, bus fares are updated every 15 minutes by the protocol issuing a transaction such as \AddrHref{https://scrollscan.com/tx/0xcced9881b3ee10fc5ece036b6a3d7df0f2101a474c8cda43206ce342173be5f9}{0xcced9881b3ee10fc5ece036b6a3d7df0f2101a474c8cda43206ce342173be5f9}).

\begin{figure}[h]
    \centering
    \includegraphics[width=0.48\textwidth]{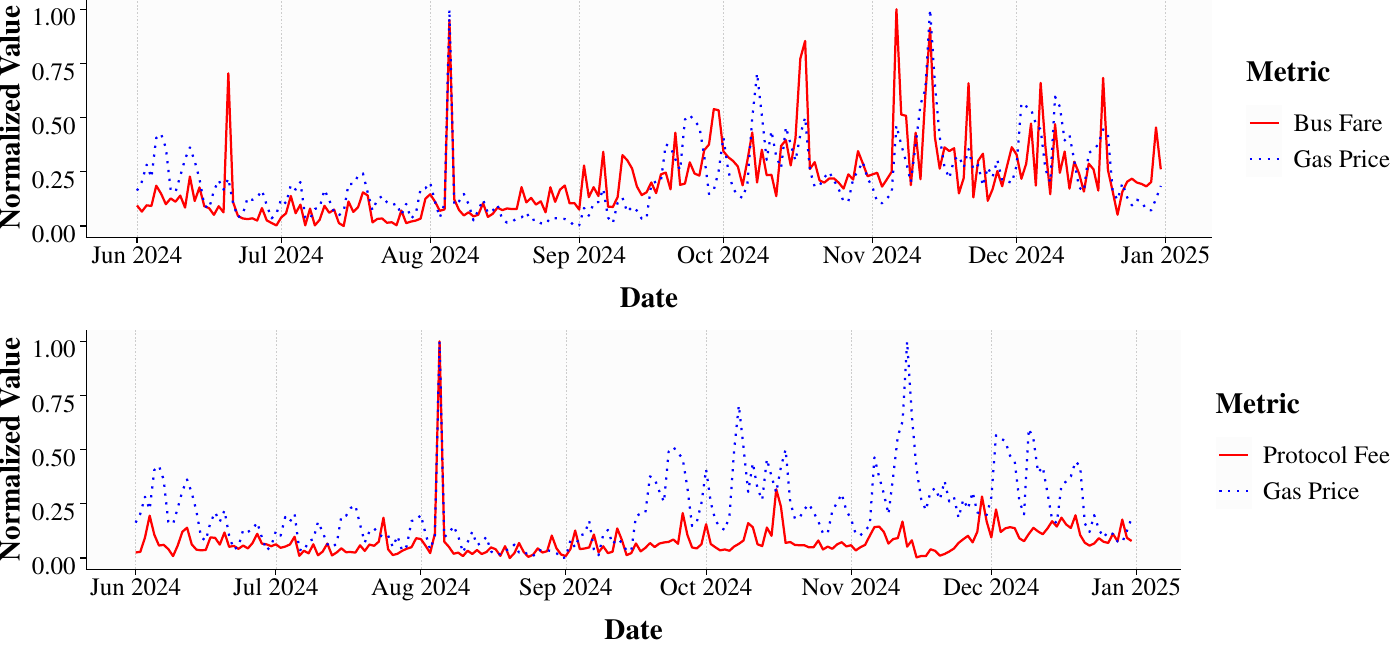} 
    \caption{Correlation of the fees paid by the user in Stargate (Bus Mode) with the Ethereum gas price (all values are normalized).}
    \label{fig: stargate_bus_fee_vs_gas_price}
    \Description[<short description>]{<long description>}
\end{figure}

\begin{figure}[h]
    \centering
    \includegraphics[width=0.48\textwidth]{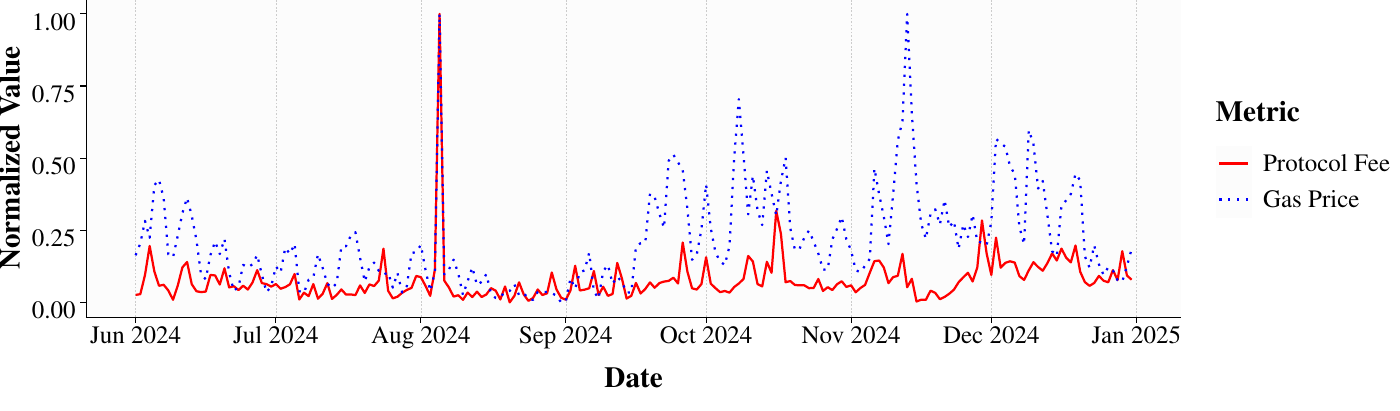} 
    \caption{Correlation of the fees paid by the user in Across with the Ethereum gas price  (all values are normalized).}
    \label{fig: across_fee_vs_gas_price}
    \Description[<short description>]{<long description>}
\end{figure}

\paragraph{Fee Models (dynamic vs. flat).}
Although there are clear trends 
in Figure~\ref{fig: arb_base_cost_per_token}, there are still multiple data points scattered throughout the plot, suggesting additional variables that impact the fees paid by the user. 
These values are computed off-chain, so we could not inspect the source code -- and later written into contracts. We hypothesized that, at each moment, the protocol fees charged were correlated with Ethereum's gas price because the cost of issuing transactions on the destination blockchains is tied to it. To analyze this correlation, Figure~\ref{fig: stargate_bus_fee_vs_gas_price} plots the bus fare and bus protocol fee paid by the user in Stargate's Bus mode (recall the fees paid in Table~\ref{table: summary}). The main difference between these fees is the way they are calculated. The bus fare is a flat fee charged by the protocol, updated every 15 minutes. The plot shows a high correlation (66\%) between the fee and the price of Ethereum gas. On the other hand, the protocol fee is calculated ``on the fly'' based on the tokens and amounts that are being transferred (according to Figure~\ref{fig: arb_base_cost_per_token}) -- these additional variables justify a lower correlation (49\%). The same findings and rationale can be drawn from Across's protocol (cf. Figure~\ref{fig: across_fee_vs_gas_price}) -- correlation of 57\% between both variables -- where the fee is also based on the token and amounts transferred.

\begin{figure}[ht]
    \centering
    \includegraphics[width=0.5\textwidth]{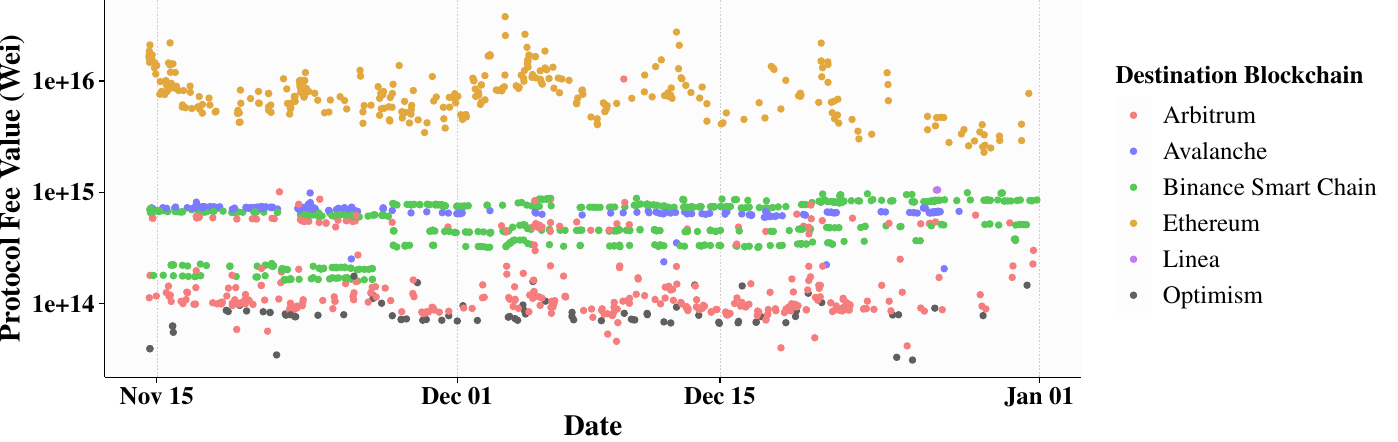}
    \caption{Protocol fee charged by CCIP when transferring Wrapped Ether (WETH) from Base to various blockchains.}
    \label{fig: ccip_fee_base_weth}
    \Description[<short description>]{<long description>}
\end{figure}

We also analyzed the CCIP fees taking as an example the transfer of WETH from Base to multiple blockchains (Figure~\ref{fig: ccip_fee_base_weth}). Here, the fee remains largely constant, with noticeable deviations, mainly when transferring to Ethereum. The same analysis done above shows that the Ethereum gas price has a similar influence on CCIP's flat fee, when transferring to Ethereum (correlation of 61\%).

\section{Discussion}
\label{sec: discussion}

In this section, we discuss our findings and summarize the answers to the different research questions posed in Section~\ref{subsec: rqs}.

\subsection{RQ1}

Firstly, we discuss \toolName's capabilities and limitations.
\toolName~supports more protocols than those analyzed in this paper (the complete list is in Section~\ref{sec: solution}) -- due to size constraints, we selected a set of bridges with different security assumptions and mixed performance and cost models. Although the specific events captured in each protocol and the specific configuration files are not present in this manuscript, they are present in the linked GitHub repository. \toolName~enables reproducible experiments and large-scale research on blockchain interoperability data.

As the main limitation, our study focused exclusively on EVM-based blockchains, excluding, for example, Solana and Sui. The identification of relevant events for each bridge was conducted by scraping data from blockchain explorers and, when available, consulting project documentation. However, not all projects provide transparent information regarding the addresses of the deployed contracts. Although we made serious efforts to capture the various state changes across blockchains, events may have been missed. Finally, a configuration file maps contract addresses to the events each bridge emits. If a protocol introduces breaking changes to its event data models, we must update the database models and listen for the new events. Since off-chain protocol components depend on these models too, such modifications are relatively infrequent.

We do not evaluate \toolName's performance because it depends on multiple factors, including the number of contracts involved in each bridge, the volume of events emitted by these contracts, and the transaction frequency per bridge (dependent on the period under analysis). Depending on these variables, the analysis duration can range from minutes to several days. However, we thoroughly tested the results and cross-referenced findings with block explorer data whenever possible. Additionally, since no other open datasets of cross-chain transactions exist, we could not compare our findings with prior results or assess potential true/false positives and negatives.

\subsection{RQ2}

This subsection reviews findings related to the design trade-offs.

\subsubsection{Latency vs. Cost vs. Security}

The solutions analyzed clearly illustrate a trade-off between security and performance. On one end of the spectrum, protocols such as CCTP and CCIP emphasize security by waiting for full finality, resulting in higher latencies. Conversely, solutions such as Stargate Taxi and Across prioritize performance, achieving lower latencies, albeit with different security guarantees. Stargate Bus mode occupies a middle ground -- it sacrifices some security guarantees by not waiting for full finality but offers reduced costs for interactions with expensive blockchains. Moreover, Across, with its cross-chain intent model (based on EIP-7683), introduces an entirely different risk profile: the risk shifts from the protocol to the solvers fulfilling user orders -- as \emph{cctxs} are optimistically accepted and only verified after the order fulfillment. More research is required on the profitability and risks faced by solvers. Across tends to be more expensive than other protocols (except for CCIP's flat fees), highlighting a trade-off, in which, to prioritize performance and overcome the security risks taken by solvers, the protocol is more expensive. Cross-chain intent-based bridges also introduce potential limitations. A primary concern is liquidity. If there is insufficient liquidity (i.e., \emph{solvers} do not have enough funds in all supported blockchains) to satisfy all user orders, the system may face execution delays or increased operating costs to maintain balanced liquidity on all blockchains.

The security and performance of a cross-chain protocol are primarily driven highly by the characteristics of the source blockchain -- e.g., in terms of finality (e.g., Section~\ref{section: l1-l1}) and block time (e.g., Section~\ref{section: l2-l1}). When the finality in the source blockchain is negligible, the architecture and proving systems used by the off-chain mechanisms play a bigger role (cf. L2 $\rightarrow$ L2 analysis in Table~\ref{table: l2-l2}). Moreover, our analysis reveals the absence of some relationships between various factors -- e.g., higher-value token transfers do not necessarily result in longer processing times (cf. Figure~\ref{fig: arb_base_cost_per_token}). Although some theoretical information exists in this direction -- e.g., assigning delays or setting withdrawal limits depending on the tiers of value transferred~\cite{ronin_withd_limits, ccip_rate_limits} -- no correlations exist in the data. 
The absence of these relationships has enabled high-profile attacks on cross-chain protocols, particularly in a vulnerable ecosystem that has suffered losses exceeding 3 billion USD since 2021~\cite{10174993}.

\subsubsection{Batching}

Batching mechanisms, such as those employed by Stargate's Bus mode, represent a trade-off between latency and user cost. Our results show that, particularly when transferring to Ethereum, the IQR for the cross-chain latency is much higher in this batching mode. This is likely due to the protocol waiting longer to fill available ``seats'' on the bus before executing the final transaction on Ethereum -- the most expensive blockchain in our dataset. Although batching can reduce the cost per transaction when the source chain is expensive -- the transaction cost is amortized by the number of cctxs to which the transaction is related -- our analysis also reveals that 42.84\% of all buses are driven with only a single user, raising questions about the overall cost-effectiveness of the protocol. However, when dealing with blockchains with low transaction fees (mainly L2s), the batching mode actually becomes more expensive (cf. Tables~\ref{table: l2-l1} and~\ref{table: l2-l2}) because the cost of issuing transactions to the blockchains becomes negligible and the user still has to pay additional fees -- namely the bus fare.

\subsection{RQ3}

Our work opens several promising directions for future research:
\begin{itemize}
    \item \textbf{In-depth Security Analysis:} As highlighted above, the cross-chain ecosystem has suffered numerous losses to attackers, exceeding 3.2 billion USD~\cite{10174993}. With our framework and datasets openly accessible, researchers can perform detailed security assessments of cross-chain protocols, including studies on previous cross-chain hacks, cross-chain MEV extraction (e.g., arbitrage -- as demonstrated in Section~\ref{section: l2-l2}), and potential price oracle manipulations -- as also demonstrated in Section~\ref{section: l2-l2}.
    \item \textbf{Native Bridges vs. General Bridges:} In this paper, we analyzed bridges on top of cross-chain communication protocols (cf. Figure~\ref{fig: amb-architecture}). However, some solutions --- called \textit{native bridges} are tailored to specific blockchain pairs, often facilitating interoperability between Ethereum and a sidechain, such as the Ronin bridge (Ethereum to Ronin) or the Polygon bridge (Ethereum to Polygon). Additionally, the results can be compared to native bridges for rollups, such as the Arbitrum bridge (Ethereum to Arbitrum) or Base ridge (Ethereum to Base). A thorough comparison between native bridges and bridges built on top of cross-chain communication protocols that support the same pair of blockchains can clarify additional trade-offs between security, performance, and cost. Although native bridges may impose waiting periods of up to seven days for L2 $\rightarrow$ L1 communication (e.g., optimistic rollups~\cite{rollups}), cross-chain protocols built on AMP protocols typically function in a range from up to $\approx$1,500 seconds -- highlighting a distinct trade-off between security assurances and \emph{cctx} speed.
    \item \textbf{Cross-Chain Activity} Another interesting dimension is the relationship between unique depositors and unique recipients across transactions (cf.\ Table~\ref{table: summary}). Initial observations indicate that users often intentionally specify a different recipient from the same depositor address on the source chain, suggesting that cross-chain transfers are not solely used for fund transfers, but may represent deliberate token mixing. A more granular analysis of these user patterns could shed light on user behavior and the underlying intent behind cross-chain activity.
    \item \textbf{Standardization of Data Models:} As outlined in Sections~\ref{subsec: extractor} and~\ref{subsec: generator}, we were required to develop multiple components (decoders and generators) for each bridge due to the unstandardized architecture of the cross-chain protocols analyzed. Developing and adopting standardized events and data models could simplify future analyses, allowing more direct comparisons between different interoperability protocols.
    \item \textbf{Analysis beyond Token Transfers:} Our framework demonstrates robust applicability for analyzing cross-chain token transfers, but it can be extended to evaluate other interoperability protocols and even non-token-related events. Arbitrary Message Passing protocols (cf. Figure~\ref{fig: amb-architecture}), at its core, allow the transfer of data between blockchains. Therefore, analyzing the messaging events captured in the \emph{cross-chain communication layer} could facilitate more detailed analyses of specific DApps or service layers built on top of these protocols.
\end{itemize}

\section{Related Work}
\label{sec: related_work}

Several efforts have been made to study cross-chain solutions from both theoretical and empirical perspectives. Some works provide a comprehensive overview of cross-chain communication protocols ~\cite{zamyatin2021sok} and general cross-chain solutions ~\cite{belchior2021survey}, while others focus on classifying these systems based on security risks and privacy guarantees ~\cite{augusto_sok_2024, Abebe_Crosschain_Risk_Framework}. Additional studies have analyzed the theoretical vulnerabilities behind cross-chain bridge attacks ~\cite{augusto_sok_2024, lee2023sok} or developed monitoring frameworks for interoperability protocols~\cite{belchior_hephaestus_2022,  zhang_xscope_2022}. On the empirical front, \cite{huang2024seamlessly} investigates cross-chain swaps on the Stargate Bridge using data collected from 2022 and 2023. Although this study offers valuable information on cross-chain dynamics, its dataset and associated analysis repository were not made public. Similarly, \cite{zhang_xscope_2022} analyzed vulnerabilities in four bridges (THORChain, pNetwork, Anyswap, and Qubit Bridge); while specific transactions highlighting vulnerabilities are publicly available, the overall dataset remains unpublished and its scale is unclear.
Although these studies are valuable, they are based on limited and outdated datasets. Our tool now enables the generation of more comprehensive and current data, thereby paving the way for further reproducible research in this area.

\section{Conclusion}
\label{sec: conclusion}

This paper addresses the pressing need for open datasets of cross-chain data to better understand protocols that move tens of billions of USD. We propose \toolName, a cross-chain dataset generation framework. We demonstrate the capabilities of this framework by collecting data from 5 cross-chain bridges and open source the first dataset of over 11 million cross-chain transactions across 11 blockchains. Based on the extracted data, we provide the first empirical comparison of cross-chain protocols, highlighting their design trade-offs between security, cost, and performance. In addition to this analysis, we identified some anomalies in cross-chain transactions, such as users incurring a profit due to the miscalculation of token prices. Moreover, this paper presents the first overview of the capabilities of cross-chain intent-based solutions, revolutionizing user experience when interacting with cross-chain protocols -- while introducing new security trade-offs and design choices. From this moment onward, the dataset and the analysis presented allow users to make wiser choices regarding the best protocols to be used under different circumstances (e.g., prioritizing security over performance). Finally, the first large-scale open dataset of cross-chain transactions opens the door to security assessments of cross-chain protocols or multiple other analyses in the DeFi ecosystem, such as cross-chain arbitrage.

\bibliographystyle{ACM-Reference-Format}
\bibliography{references.bib}




\section{Performance and Cost Results}\label{appendix: performance_and_cost_results}

\input{tables/results-latency.tex}

\input{tables/results-cost.tex}

\end{document}

%% file: tables/summary.tex
\setlength{\tabcolsep}{3pt} 

\begin{table}[t]
\begin{threeparttable}
\scriptsize
\begin{tabular}{lrrrrrr}
\toprule
 & \textbf{CCTP} & \textbf{CCIP} & \textbf{\begin{tabular}[c]{@{}c@{}}Stargate\\(Taxi)\end{tabular}} & \textbf{\begin{tabular}[c]{@{}c@{}}Stargate\\(Bus)\end{tabular}} & \textbf{Across} & \textbf{Total} \\
\midrule
Num of \emph{cctxs} & 592,141 & 11,430 & 3,291,711 & 3,526,050 & 3,864,421 & 11,285,753 \\
Num of Unique Chains & 6 & 9 & 9 & 9 & 7 & 10 \\
\addlinespace[0.05cm]
\hdashline
\addlinespace[0.08cm]
Total Value (\$ M) & 8,567.50 & 427.71 & 10,413.35 & 3,275.53 & 5,513.69 & 28,197.79 \\
Q1 (25th Percentile \$) & 10.73 & 350.99 & 34.80 & 18.62 & 10.25 & 17.37 \\
Q2 (50th Percentile \$) & 150.39 & 1,665.12 & 154.02 & 43.89 & 51.95 & 64.39 \\
Q3 (75th Percentile \$) & 1,199.97 & 19,971.28 & 574.99 & 252.08 & 294.96 & 383.45 \\
Max (\$ M) & 1.00 & 9.72 & 287.39 & 5.00 & 3.57 & 287.39 \\
\addlinespace[0.05cm]
\hdashline
\addlinespace[0.08cm]
Unique Tokens & 1 & 71 & 5 & 4 & 104 & 169 \\
Unique Depositors & 16,291 & 1,749 & 215,706 & 722,251 & 737,176 & 916,405 \\
Unique Recipients & 121,924 & 2,228 & 592,992 & 722,251 & 1,079,408 & 1,765,837 \\
\addlinespace[0.08cm]
\hdashline
\addlinespace[0.1cm]
User Paid Fees & \multicolumn{1}{c}{--} & \multicolumn{1}{c}{Flat Fee\tnote{\textdagger}} & \multicolumn{1}{c}{--} & \resizebox{0.4cm}{!}{$q_{a_{B_1}}$} -- \resizebox{0.4cm}{!}{$q_{a_{B_2}}$\tnote{\textdaggerdbl}} & \resizebox{0.4cm}{!}{$q_{a_{B_1}}$} -- \resizebox{0.4cm}{!}{$q_{a_{B_2}}$\tnote{\textdaggerdbl}} & \multicolumn{1}{c}{--} \\
\addlinespace[0.05cm]
& &  &  & \multicolumn{1}{c}{+ \;Flat Fee\tnote{\textdagger*}} &  &  \\
\bottomrule
\end{tabular}
\begin{tablenotes}
   \item [\textdagger] A flat (i.e., constant) configured by the protocol with a certain periodicity through a transaction.
   \item [\textdaggerdbl] the fee charged by the protocol is obtained by subtracting the input amount with the output amount (when both amounts are normalized -- e.g., in USD)
\end{tablenotes}
\end{threeparttable}

\caption{Summary of the data generated for each bridge during the last 6 months of 2024 (Jun 1, 2024 to Dec 31, 2024)}
\label{table: summary}
\end{table}

\tnote{*}

%% file: tables/L1-L1-latency-cost.tex
\setlength{\tabcolsep}{4pt}

\begin{table}[ht]
\rowcolors{2}{gray!10}{white}
\tiny
\begin{tabular}{lrrrrrrrr}
\toprule
\textbf{L1 $\rightarrow$ L1}  & \multicolumn{2}{c}{\textbf{CCTP}} & \multicolumn{2}{c}{\textbf{CCIP}} & \multicolumn{2}{c}{\textbf{Stargate (Taxi)}} & \multicolumn{2}{c}{\textbf{Stargate (Bus)}} \\
\cmidrule(r){2-3} \cmidrule(rl){4-5} \cmidrule(rl){6-7} \cmidrule(l){8-9}
 & E $\rightarrow$ A & A $\rightarrow$ E & E $\rightarrow$ A & A $\rightarrow$ E & E $\rightarrow$ A & A $\rightarrow$ E & E $\rightarrow$ A & A $\rightarrow$ E \\
\hline
Num CCTXs & 2211 & 4126 & 113 & 71 & 838 & 704 & 374 & 1085 \\
Value Trsf (\$) & \$ 139.80M & \$ 148.72M & \$ 12.12M & \$ 8.98M & \$ 7.28M & \$ 13.17M & \$ 3.66M & \$ 5.82M \\
\hline
\multicolumn{9}{c}{\textbf{Latency (s)}} \\
Q1 & 926.00 & 32.00 & 915.00 & 103.50 & 209.00 & 132.00 & 452.00 & 352.00 \\
Q2 & \cellcolor{red!25}1,044.00 & \cellcolor{blue!25}57.00 & \cellcolor{red!25}983.00 & \cellcolor{blue!25}138.00 & \cellcolor{red!25}212.00 & \cellcolor{blue!25}139.00 & \cellcolor{red!25}464.00 & \cellcolor{blue!25}632.00 \\
Q3 & 1,147.00 & 92.00 & 1,085.00 & 168.00 & 223.00 & 149.00 & 482.75 & 670.00 \\
IQR & 221.00 & 60.00 & 170.00 & 64.50 & 14.00 & 17.00 & 30.75 & 318.00 \\
\hline
\multicolumn{9}{c}{\textbf{Cost (\$)}} \\
Q1 & 2.62 & 0.12 & 11.24 & 18.15 & 7.91 & 0.20 & 2.29 & 1.62 \\
Q2 & \cellcolor{red!25}5.56 & \cellcolor{blue!25}0.20 & \cellcolor{red!25}14.87 & \cellcolor{blue!25}23.26 & \cellcolor{red!25}15.23 & \cellcolor{blue!25}0.35 & \cellcolor{red!25}4.47 & \cellcolor{blue!25}4.50 \\
Q3 & 11.25 & 0.33 & 17.51 & 30.50 & 24.87 & 0.56 & 9.42 & 13.05 \\
IQR & 8.63 & 0.21 & 6.27 & 12.36 & 16.97 & 0.37 & 7.13 & 11.43 \\
\bottomrule
\end{tabular}
\caption{Quartiles of the distribution of latencies and costs of \emph{cctxs} between Ethereum (E) and Avalanche (A), between Jun 1, 2024 and Jan 1, 2025. The color coding is set to help comparing values in the same direction (E $\rightarrow$ A or A $\rightarrow$ E).}. 
\label{table: l1-l1}
\end{table}

%% file: tables/L1-L2-latency-cost.tex
\begin{table}[ht]
\rowcolors{2}{gray!10}{white}
\tiny
\setlength{\tabcolsep}{2.5pt}
\begin{tabular}{lcccccccccc}
\toprule
\textbf{L1 $\rightarrow$ L2}  & \multicolumn{2}{c}{\textbf{CCTP}} & \multicolumn{2}{c}{\textbf{CCIP}} & \multicolumn{2}{c}{\textbf{Stargate (Taxi)}} & \multicolumn{2}{c}{\textbf{Stargate (Bus)}} & \multicolumn{2}{c}{\textbf{Across}} \\
\cmidrule(r){2-3} \cmidrule(rl){4-5} \cmidrule(rl){6-7} \cmidrule(l){8-9} \cmidrule(l){10-11}
 & E $\rightarrow$ B & A $\rightarrow$ B & E $\rightarrow$ B & A $\rightarrow$ B & E $\rightarrow$ B & A $\rightarrow$ B & E $\rightarrow$ B & A $\rightarrow$ B & E $\rightarrow$ B & A $\rightarrow$ B \\
\hline
Num CCTXs & 23882 & 14238 & 461 & 126 & 13487 & 4421 & 16461 & 3313 & 279910 &  \\
Value Trsf (\$) & 837.29M & 203.42M & 8.51M & 10.35M & 163.79M & 4.29M & 80.02M & 1.48M & 531.03M &  \\
\hline
\multicolumn{11}{c}{\textbf{Latency (s)}} \\
Q1 & 936.00 & 22.00 & 964.00 & 61.25 & 210.00 & 52.00 & 436.00 & 247.00 & 16.00 &  \\
Q2 & \cellcolor{red!25}1,064.00 & \cellcolor{blue!25}63.00 & \cellcolor{red!25}1,060.00 & \cellcolor{blue!25}77.00 & \cellcolor{red!25}212.00 & \cellcolor{blue!25}54.00 & \cellcolor{red!25}458.00 & \cellcolor{blue!25}253.00 & \cellcolor{red!25}18.00 &  \\
Q3 & 1,204.00 & 103.00 & 1,160.00 & 93.00 & 224.00 & 57.00 & 508.00 & 367.00 & 28.00 &  \\
IQR & 268.00 & 81.00 & 196.00 & 31.75 & 14.00 & 5.00 & 72.00 & 120.00 & 12.00 &  \\
\hline
\multicolumn{11}{c}{\textbf{Cost (\$)}} \\
Q1 & 1.51 & 0.13 & 9.32 & 1.80 & 5.68 & 0.18 & 1.25 & 0.16 & 2.04 &  \\
Q2 & \cellcolor{red!25}3.51 & \cellcolor{blue!25}0.23 & \cellcolor{red!25}12.87 & \cellcolor{blue!25}2.11 & \cellcolor{red!25}11.04 & \cellcolor{blue!25}0.26 & \cellcolor{red!25}2.56 & \cellcolor{blue!25}0.25 & \cellcolor{red!25}3.47 &  \\
Q3 & 7.73 & 0.39 & 18.71 & 2.29 & 19.64 & 0.48 & 4.88 & 0.41 & 5.80 &  \\
IQR & 6.22 & 0.26 & 9.39 & 0.49 & 13.96 & 0.30 & 3.63 & 0.25 & 3.76 &  \\
\bottomrule
\end{tabular}
\caption{Quartiles of the distribution of latencies and costs of \emph{cctxs} from Ethereum (E) to Base (B), and Avalanche (A) to Base (B), between Jun 1, 2024 and Jan 1, 2025.}.
\label{table: l1-l2}
\end{table}

%% file: tables/L2-L1-latency-cost.tex
\begin{table}[h]
\setlength{\tabcolsep}{2.5pt}
\rowcolors{2}{gray!10}{white}
\tiny
\begin{tabular}{lcccccccccc}
\toprule
\textbf{L2 $\rightarrow$ L1}  & \multicolumn{2}{c}{\textbf{CCTP}} & \multicolumn{2}{c}{\textbf{CCIP}} & \multicolumn{2}{c}{\textbf{Stargate (Taxi)}} & \multicolumn{2}{c}{\textbf{Stargate (Bus)}} & \multicolumn{2}{c}{\textbf{Across}} \\
\cmidrule(r){2-3} \cmidrule(rl){4-5} \cmidrule(rl){6-7} \cmidrule(l){8-9} \cmidrule(l){10-11}
 & B $\rightarrow$ E & B $\rightarrow$ A & B $\rightarrow$ E & B $\rightarrow$ A & B $\rightarrow$ E & B $\rightarrow$ A & B $\rightarrow$ E & B $\rightarrow$ A & B $\rightarrow$ E & B $\rightarrow$ A \\
\hline
Num CCTXs & 22272 & 8598 & 353 & 308 & 7493 & 4023 & 56949 & 2326 & 140641 &  \\
Value Trsf (\$) & 616.46M & 124.99M & 8.54M & 17.09M & 216.20M & 6.54M & 84.22M & 2.56M & 669.63M &  \\
\hline
\multicolumn{11}{c}{\textbf{Latency (s)}} \\
Q1 & 1,198.00 & 1,209.00 & 1,334.00 & 1,257.75 & 126.00 & 46.00 & 228.00 & 241.00 & 4.00 &  \\
Q2 & \cellcolor{red!25}1,328.00 & \cellcolor{blue!25}1,326.00 & \cellcolor{red!25}1,444.00 & \cellcolor{blue!25}1,338.00 & \cellcolor{red!25}132.00 & \cellcolor{blue!25}48.00 & \cellcolor{red!25}340.00 & \cellcolor{blue!25}246.00 & \cellcolor{red!25}8.00 &  \\
Q3 & 1,510.00 & 1,452.00 & 1,526.00 & 1,448.25 & 144.00 & 51.00 & 592.00 & 362.00 & 12.00 &  \\
IQR & 312.00 & 243.00 & 192.00 & 190.50 & 18.00 & 5.00 & 364.00 & 121.00 & 8.00 &  \\
\hline
\multicolumn{11}{c}{\textbf{Cost (\$)}} \\
Q1 & 0.00 & 0.00 & 19.42 & 2.32 & 0.00 & 0.01 & 0.62 & 0.16 & 3.95 &  \\
Q2 & \cellcolor{red!25}0.00 & \cellcolor{blue!25}0.01 & \cellcolor{red!25}26.64 & \cellcolor{blue!25}2.38 & \cellcolor{red!25}0.01 & \cellcolor{blue!25}0.01 & \cellcolor{red!25}1.54 & \cellcolor{blue!25}0.28 & \cellcolor{red!25}6.71 &  \\
Q3 & 0.01 & 0.02 & 36.63 & 2.45 & 0.02 & 0.02 & 3.76 & 0.87 & 10.72 &  \\
IQR & 0.01 & 0.01 & 17.21 & 0.13 & 0.01 & 0.02 & 3.13 & 0.71 & 6.77 &  \\
\bottomrule
\end{tabular}
\caption{Quartiles of the distribution of latencies and costs of \emph{cctxs} from Base (B) to Ethereum (E), and Base (B) to Avalanche (A), between Jun 1, 2024 and Jan 1, 2025.}.
\label{table: l2-l1}
\end{table}

%% file: tables/L2-L2-latency-cost.tex
\setlength{\tabcolsep}{2pt}

\begin{table}[ht]
\rowcolors{2}{gray!10}{white}
\tiny
\begin{tabular}{lcccccccccc}
\toprule
\textbf{L2 $\rightarrow$ L2}  & \multicolumn{2}{c}{\textbf{CCTP}} & \multicolumn{2}{c}{\textbf{CCIP}} & \multicolumn{2}{c}{\textbf{Stargate (Taxi)}} & \multicolumn{2}{c}{\textbf{Stargate (Bus)}} & \multicolumn{2}{c}{\textbf{Across}} \\
\cmidrule(r){2-3} \cmidrule(rl){4-5} \cmidrule(rl){6-7} \cmidrule(l){8-9} \cmidrule(l){10-11}
 & B $\rightarrow$ A & A $\rightarrow$ B & B $\rightarrow$ A & A $\rightarrow$ B & B $\rightarrow$ A & A $\rightarrow$ B & B $\rightarrow$ A & A $\rightarrow$ B & B $\rightarrow$ A & A $\rightarrow$ B \\
\hline
Num CCTXs & 71532 & 76498 & 379 & 450 & 125215 & 287908 & 198483 & 208515 & 278554 & 439916 \\
Value Trsf (\$) & 311.32M & 288.61M & 4.88M & 5.87M & 306.02M & 421.69M & 174.71M & 170.44M & 203.03M & 239.04M \\
\hline
\multicolumn{11}{c}{\textbf{Latency (s)}} \\
Q1 & 1,228.00 & 1,013.00 & 1,263.50 & 1,029.00 & 29.00 & 31.00 & 61.00 & 57.00 & 1.00 & 2.00 \\
Q2 & \cellcolor{red!25}1,347.00 & \cellcolor{blue!25}1,119.00 & \cellcolor{red!25}1,378.00 & \cellcolor{blue!25}1,130.50 & \cellcolor{red!25}30.00 & \cellcolor{blue!25}36.00 & \cellcolor{red!25}109.00 & \cellcolor{blue!25}101.00 & \cellcolor{red!25}1.00 & \cellcolor{blue!25}3.00 \\
Q3 & 1,480.00 & 1,218.00 & 1,487.50 & 1,232.00 & 34.00 & 46.00 & 203.00 & 198.00 & 2.00 & 3.00 \\
IQR & 252.00 & 205.00 & 224.00 & 203.00 & 5.00 & 15.00 & 142.00 & 141.00 & 1.00 & 1.00 \\
\hline
\multicolumn{11}{c}{\textbf{Cost (\$)}} \\
Q1 & 0.01 & 0.02 & 0.35 & 0.34 & 0.00 & 0.01 & 0.03 & 0.02 & 0.02 & 0.01 \\
Q2 & \cellcolor{red!25}0.01 & \cellcolor{blue!25}0.03 & \cellcolor{red!25}0.45 & \cellcolor{blue!25}0.44 & \cellcolor{red!25}0.01 & \cellcolor{blue!25}0.01 & \cellcolor{red!25}0.03 & \cellcolor{blue!25}0.03 & \cellcolor{red!25}0.04 & \cellcolor{blue!25}0.02 \\
Q3 & 0.02 & 0.04 & 1.79 & 1.00 & 0.02 & 0.03 & 0.06 & 0.06 & 0.08 & 0.05 \\
IQR & 0.02 & 0.02 & 1.44 & 0.66 & 0.02 & 0.02 & 0.03 & 0.03 & 0.06 & 0.04 \\
\bottomrule
\end{tabular}
\caption{Quartiles of the distribution of latencies and costs of \emph{cctxs} between Base (B), and Arbitrum (A), between Jun 1, 2024 and Jan 1, 2025.}.
\label{table: l2-l2}
\end{table}

%% file: tables/results-latency.tex
\setlength{\tabcolsep}{4pt} 

\begin{landscape}

\begin{table}[ht]
\rowcolors{2}{gray!10}{white}
\tiny
\begin{tabular}{lrrrrrrrrrrrrrrrrrrrrrrrrr}
\toprule
 & \multicolumn{5}{c}{\textbf{cctp}} & \multicolumn{5}{c}{\textbf{ccip}} & \multicolumn{5}{c}{\textbf{stargate\_oft}} & \multicolumn{5}{c}{\textbf{stargate\_bus}} & \multicolumn{5}{c}{\textbf{across}} \\
\cmidrule(rl){2-6} \cmidrule(rl){7-11} \cmidrule(rl){12-16} \cmidrule(rl){17-21} \cmidrule(rl){22-26} 
 & \multicolumn{1}{c}{$n$} & \multicolumn{1}{c}{Q1} & \multicolumn{1}{c}{Q2} & \multicolumn{1}{c}{Q3} & \multicolumn{1}{c}{IQR} & \multicolumn{1}{c}{$n$} & \multicolumn{1}{c}{Q1} & \multicolumn{1}{c}{Q2} & \multicolumn{1}{c}{Q3} & \multicolumn{1}{c}{IQR} & \multicolumn{1}{c}{$n$} & \multicolumn{1}{c}{Q1} & \multicolumn{1}{c}{Q2} & \multicolumn{1}{c}{Q3} & \multicolumn{1}{c}{IQR} & \multicolumn{1}{c}{$n$} & \multicolumn{1}{c}{Q1} & \multicolumn{1}{c}{Q2} & \multicolumn{1}{c}{Q3} & \multicolumn{1}{c}{IQR} & \multicolumn{1}{c}{$n$} & \multicolumn{1}{c}{Q1} & \multicolumn{1}{c}{Q2} & \multicolumn{1}{c}{Q3} & \multicolumn{1}{c}{IQR} \\
\hline
ethereum $\rightarrow$ avalanche & 2211 & 926.00 & 1044.00 & 1147.00 & 221.00 & 113 & 915.00 & 983.00 & 1085.00 & 170.00 & 1676 & 209.00 & 212.00 & 223.00 & 14.00 & 374 & 452.00 & 464.00 & 482.75 & 30.75 & & & & &  \\
ethereum $\rightarrow$ bnb & & & & &  & 591 & 939.00 & 1041.00 & 1137.50 & 198.50 & 9129 & 214.00 & 218.00 & 229.00 & 15.00 & 2945 & 455.00 & 464.00 & 499.00 & 44.00 & & & & &  \\
avalanche $\rightarrow$ ethereum & 4126 & 32.00 & 57.00 & 92.00 & 60.00 & 71 & 103.50 & 138.00 & 168.00 & 64.50 & 1408 & 132.00 & 139.00 & 149.00 & 17.00 & 1085 & 352.00 & 632.00 & 670.00 & 318.00 & & & & &  \\
avalanche $\rightarrow$ bnb & & & & &  & 303 & 56.00 & 74.00 & 88.50 & 32.50 & 6279 & 56.00 & 59.00 & 63.00 & 7.00 & 4219 & 251.00 & 267.00 & 372.00 & 121.00 & & & & &  \\
bnb $\rightarrow$ avalanche & & & & &  & 457 & 66.00 & 80.00 & 97.00 & 31.00 & 6896 & 86.00 & 88.00 & 91.00 & 5.00 & 2726 & 283.00 & 288.00 & 368.00 & 85.00 & & & & &  \\
bnb $\rightarrow$ ethereum & & & & &  & 286 & 129.00 & 154.00 & 184.75 & 55.75 & 13327 & 166.00 & 172.00 & 182.00 & 16.00 & 8684 & 404.75 & 665.00 & 682.00 & 277.25 & & & & &  \\
\hspace*{0.5cm}\textbf{L1 $\rightarrow$ L1} & 6337 & 46.00 & 97.00 & 953.00 & 907.00 & 1821 & 82.00 & 154.00 & 979.00 & 897.00 & 38715 & 88.00 & 169.00 & 212.00 & 124.00 & 20033 & 296.00 & 440.00 & 663.00 & 367.00 & & & & &  \\
\hline
ethereum $\rightarrow$ base & 23882 & 936.00 & 1064.00 & 1204.00 & 268.00 & 461 & 964.00 & 1060.00 & 1160.00 & 196.00 & 27002 & 210.00 & 212.00 & 224.00 & 14.00 & 16476 & 436.00 & 458.00 & 508.00 & 72.00 & 279910 & 16.00 & 18.00 & 28.00 & 12.00 \\
ethereum $\rightarrow$ arbitrum & 17135 & 977.00 & 1092.00 & 1247.00 & 270.00 & 443 & 951.00 & 1042.00 & 1125.50 & 174.50 & 22117 & 192.00 & 203.00 & 217.00 & 25.00 & 20554 & 399.00 & 437.00 & 470.00 & 71.00 & 94264 & 2.00 & 26.00 & 31.00 & 29.00 \\
ethereum $\rightarrow$ optimism & 3964 & 990.00 & 1120.00 & 1264.00 & 274.00 & 22 & 957.00 & 1065.00 & 1186.50 & 229.50 & 3656 & 208.00 & 220.00 & 226.00 & 18.00 & 3635 & 448.00 & 462.00 & 480.00 & 32.00 & 23022 & 10.00 & 28.00 & 32.00 & 22.00 \\
ethereum $\rightarrow$ linea & & & & &  & 2 & 3412.50 & 5820.00 & 8227.50 & 4815.00 & 2214 & 200.25 & 212.00 & 227.00 & 26.75 & 8017 & 363.00 & 440.00 & 466.00 & 103.00 & 25856 & 17.00 & 29.00 & 31.00 & 14.00 \\
ethereum $\rightarrow$ scroll & & & & &  & & & & &  & 1647 & 227.00 & 244.00 & 254.00 & 27.00 & 6078 & 455.00 & 476.00 & 502.00 & 47.00 & 4467 & 9.00 & 22.00 & 28.00 & 19.00 \\
avalanche $\rightarrow$ base & 14238 & 22.00 & 63.00 & 103.00 & 81.00 & 126 & 61.25 & 77.00 & 93.00 & 31.75 & 8870 & 52.00 & 54.00 & 57.00 & 5.00 & 3349 & 247.00 & 253.00 & 368.00 & 121.00 & & & & &  \\
avalanche $\rightarrow$ arbitrum & 18619 & 19.00 & 55.00 & 96.00 & 77.00 & 110 & 56.00 & 70.00 & 81.50 & 25.50 & 15122 & 35.00 & 37.00 & 45.00 & 10.00 & 9411 & 223.50 & 239.00 & 258.00 & 34.50 & & & & &  \\
avalanche $\rightarrow$ optimism & 5401 & 17.00 & 46.00 & 86.00 & 69.00 & 10 & 40.00 & 64.50 & 81.75 & 41.75 & 4856 & 51.00 & 53.00 & 56.00 & 5.00 & 2406 & 247.00 & 253.00 & 265.00 & 18.00 & & & & &  \\
avalanche $\rightarrow$ linea & & & & &  & 2 & 2206.25 & 4199.50 & 6192.75 & 3986.50 & & & & &  & & & & &  & & & & &  \\
avalanche $\rightarrow$ scroll & & & & &  & & & & &  & 536 & 60.00 & 69.00 & 125.25 & 65.25 & 1009 & 257.00 & 271.00 & 371.00 & 114.00 & & & & &  \\
bnb $\rightarrow$ base & & & & &  & 2060 & 68.00 & 86.00 & 108.00 & 40.00 & 38792 & 86.00 & 88.00 & 90.00 & 4.00 & 14569 & 263.00 & 284.00 & 401.00 & 138.00 & & & & &  \\
bnb $\rightarrow$ arbitrum & & & & &  & 358 & 58.00 & 74.00 & 89.00 & 31.00 & 43525 & 69.00 & 70.00 & 76.00 & 7.00 & 31172 & 167.00 & 265.00 & 284.00 & 117.00 & & & & &  \\
bnb $\rightarrow$ optimism & & & & &  & 2 & 78.75 & 83.50 & 88.25 & 9.50 & 13400 & 85.00 & 87.00 & 89.00 & 4.00 & 6966 & 281.00 & 287.00 & 400.00 & 119.00 & & & & &  \\
bnb $\rightarrow$ linea & & & & &  & 2 & 201.50 & 213.00 & 224.50 & 23.00 & & & & &  & & & & &  & & & & &  \\
bnb $\rightarrow$ scroll & & & & &  & & & & &  & 955 & 92.00 & 123.00 & 236.00 & 144.00 & 2305 & 288.00 & 313.00 & 430.00 & 142.00 & & & & &  \\
\hspace*{0.5cm}\textbf{L1 $\rightarrow$ L2} & 83239 & 63.00 & 873.00 & 1102.00 & 1039.00 & 3598 & 71.00 & 96.00 & 848.75 & 777.75 & 182692 & 70.00 & 87.00 & 206.00 & 136.00 & 125947 & 265.00 & 337.00 & 450.00 & 185.00 & 427519 & 12.00 & 18.00 & 30.00 & 18.00 \\
\hline
arbitrum $\rightarrow$ avalanche & 9209 & 994.00 & 1098.00 & 1192.00 & 198.00 & 305 & 1027.00 & 1129.00 & 1213.00 & 186.00 & 50579 & 33.00 & 40.00 & 49.00 & 16.00 & 5834 & 218.00 & 233.00 & 335.00 & 117.00 & & & & &  \\
arbitrum $\rightarrow$ ethereum & 13679 & 1059.00 & 1184.00 & 1307.00 & 248.00 & 363 & 1090.00 & 1183.00 & 1296.00 & 206.00 & 46740 & 114.00 & 126.00 & 141.00 & 27.00 & 69393 & 208.00 & 313.00 & 507.00 & 299.00 & 84100 & 5.00 & 9.00 & 15.00 & 10.00 \\
arbitrum $\rightarrow$ bnb & & & & &  & 1041 & 1020.00 & 1112.00 & 1219.00 & 199.00 & 155311 & 39.00 & 52.00 & 66.00 & 27.00 & 21385 & 199.00 & 237.00 & 263.00 & 64.00 & & & & &  \\
optimism $\rightarrow$ avalanche & 3819 & 1325.00 & 1455.00 & 1614.00 & 289.00 & 3 & 1593.00 & 1622.00 & 1653.50 & 60.50 & 5487 & 66.00 & 68.00 & 70.00 & 4.00 & 2173 & 260.00 & 266.00 & 380.00 & 120.00 & & & & &  \\
optimism $\rightarrow$ ethereum & 3267 & 1386.00 & 1548.00 & 1806.00 & 420.00 & 18 & 1578.50 & 1706.00 & 1906.00 & 327.50 & 4406 & 146.00 & 152.00 & 164.00 & 18.00 & 25077 & 300.00 & 364.00 & 640.00 & 340.00 & 23771 & 6.00 & 10.00 & 18.00 & 12.00 \\
optimism $\rightarrow$ bnb & & & & &  & 1 & 1593.00 & 1593.00 & 1593.00 & 0.00 & 11765 & 70.00 & 73.00 & 80.00 & 10.00 & 8578 & 263.00 & 268.00 & 384.00 & 121.00 & & & & &  \\
base $\rightarrow$ avalanche & 8598 & 1209.00 & 1326.00 & 1452.00 & 243.00 & 308 & 1257.75 & 1338.00 & 1448.25 & 190.50 & 8046 & 46.00 & 48.00 & 51.00 & 5.00 & 2326 & 241.00 & 246.00 & 362.00 & 121.00 & & & & &  \\
base $\rightarrow$ ethereum & 22272 & 1198.00 & 1328.00 & 1510.00 & 312.00 & 353 & 1334.00 & 1444.00 & 1526.00 & 192.00 & 14987 & 126.00 & 132.00 & 144.00 & 18.00 & 56950 & 228.00 & 340.00 & 592.00 & 364.00 & 140641 & 4.00 & 8.00 & 12.00 & 8.00 \\
base $\rightarrow$ bnb & & & & &  & 676 & 1274.00 & 1354.00 & 1456.00 & 182.00 & 26418 & 51.00 & 53.00 & 55.00 & 4.00 & 10240 & 242.00 & 247.00 & 364.00 & 122.00 & & & & &  \\
linea $\rightarrow$ avalanche & & & & &  & 2 & 1341.50 & 1343.00 & 1344.50 & 3.00 & & & & &  & & & & &  & & & & &  \\
linea $\rightarrow$ ethereum & & & & &  & 2 & 1408.75 & 1428.50 & 1448.25 & 39.50 & 4279 & 130.00 & 143.00 & 225.00 & 95.00 & 67065 & 348.00 & 441.00 & 628.00 & 280.00 & 17289 & 5.00 & 9.00 & 14.00 & 9.00 \\
linea $\rightarrow$ bnb & & & & &  & 2 & 1341.75 & 1348.50 & 1355.25 & 13.50 & & & & &  & & & & &  & & & & &  \\
scroll $\rightarrow$ avalanche & & & & &  & & & & &  & 649 & 53.00 & 89.00 & 92.00 & 39.00 & 855 & 252.00 & 290.00 & 359.00 & 107.00 & & & & &  \\
scroll $\rightarrow$ ethereum & & & & &  & & & & &  & 2894 & 133.00 & 164.00 & 179.00 & 46.00 & 27824 & 318.00 & 428.00 & 662.00 & 344.00 & 7679 & 8.00 & 12.00 & 19.00 & 11.00 \\
scroll $\rightarrow$ bnb & & & & &  & & & & &  & 1961 & 93.00 & 95.00 & 98.00 & 5.00 & 3745 & 290.00 & 296.00 & 370.00 & 80.00 & & & & &  \\
\hspace*{0.5cm}\textbf{L2 $\rightarrow$ L1} & 60844 & 1130.00 & 1267.00 & 1461.00 & 331.00 & 3074 & 1111.00 & 1240.50 & 1367.75 & 256.75 & 333522 & 45.00 & 56.00 & 110.00 & 65.00 & 301445 & 243.00 & 344.00 & 569.00 & 326.00 & 273480 & 5.00 & 8.00 & 12.00 & 7.00 \\
\hline
arbitrum $\rightarrow$ base & 76498 & 1013.00 & 1119.00 & 1218.00 & 205.00 & 450 & 1029.00 & 1130.50 & 1232.00 & 203.00 & 576116 & 31.00 & 36.00 & 46.00 & 15.00 & 208847 & 57.00 & 101.00 & 198.00 & 141.00 & 439916 & 2.00 & 3.00 & 3.00 & 1.00 \\
arbitrum $\rightarrow$ optimism & 24770 & 1009.00 & 1116.00 & 1216.00 & 207.00 & 28 & 1027.50 & 1096.50 & 1177.50 & 150.00 & 411998 & 30.00 & 36.00 & 45.00 & 15.00 & 133141 & 63.00 & 128.00 & 219.00 & 156.00 & 141592 & 2.00 & 3.00 & 4.00 & 2.00 \\
arbitrum $\rightarrow$ linea & & & & &  & 4 & 1155.50 & 1233.00 & 3046.50 & 1891.00 & 36935 & 20.00 & 23.00 & 34.00 & 14.00 & 69355 & 59.00 & 162.00 & 221.00 & 162.00 & 197122 & 2.00 & 3.00 & 12.00 & 10.00 \\
arbitrum $\rightarrow$ scroll & & & & &  & & & & &  & 24070 & 44.00 & 53.00 & 71.00 & 27.00 & 110142 & 73.00 & 166.00 & 244.00 & 171.00 & 43424 & -1.00 & 0.00 & 1.00 & 2.00 \\
optimism $\rightarrow$ base & 38762 & 1352.00 & 1496.00 & 1656.00 & 304.00 & 68 & 1505.00 & 1661.00 & 1792.00 & 287.00 & 181804 & 66.00 & 68.00 & 70.00 & 4.00 & 175736 & 96.00 & 138.00 & 238.00 & 142.00 & 264441 & 2.00 & 2.00 & 4.00 & 2.00 \\
optimism $\rightarrow$ arbitrum & 35364 & 1346.00 & 1482.00 & 1639.00 & 293.00 & 23 & 1541.00 & 1634.00 & 1803.50 & 262.50 & 170302 & 49.00 & 50.00 & 55.00 & 6.00 & 167702 & 83.00 & 138.00 & 241.00 & 158.00 & 161243 & 0.00 & 1.00 & 2.00 & 2.00 \\
optimism $\rightarrow$ linea & & & & &  & & & & &  & 22221 & 55.00 & 58.00 & 64.00 & 9.00 & 53572 & 95.00 & 207.00 & 257.00 & 162.00 & 147027 & 3.00 & 4.00 & 20.00 & 17.00 \\
optimism $\rightarrow$ scroll & & & & &  & & & & &  & 11776 & 74.00 & 82.00 & 93.00 & 19.00 & 77823 & 112.00 & 219.00 & 287.00 & 175.00 & 25391 & -1.00 & 0.00 & 1.00 & 2.00 \\
base $\rightarrow$ arbitrum & 71532 & 1228.00 & 1347.00 & 1480.00 & 252.00 & 379 & 1263.50 & 1378.00 & 1487.50 & 224.00 & 250438 & 29.00 & 30.00 & 34.00 & 5.00 & 198497 & 61.00 & 109.00 & 203.00 & 142.00 & 278554 & 1.00 & 1.00 & 2.00 & 1.00 \\
base $\rightarrow$ optimism & 37235 & 1232.00 & 1348.00 & 1474.00 & 242.00 & 84 & 1275.00 & 1376.00 & 1464.00 & 189.00 & 122690 & 46.00 & 46.00 & 48.00 & 2.00 & 139906 & 76.00 & 124.00 & 234.00 & 158.00 & 199692 & 2.00 & 2.00 & 4.00 & 2.00 \\
base $\rightarrow$ linea & & & & &  & 2 & 3369.50 & 4999.00 & 6628.50 & 3259.00 & 55973 & 35.00 & 37.00 & 42.00 & 7.00 & 87467 & 69.00 & 153.00 & 231.00 & 162.00 & 170460 & 2.00 & 3.00 & 8.00 & 6.00 \\
base $\rightarrow$ scroll & & & & &  & & & & &  & 22668 & 54.00 & 62.00 & 111.00 & 57.00 & 102918 & 95.00 & 194.00 & 264.00 & 169.00 & 50025 & 0.00 & 1.00 & 2.00 & 2.00 \\
linea $\rightarrow$ base & & & & &  & 2 & 1289.50 & 1292.00 & 1294.50 & 5.00 & 194370 & 47.00 & 50.00 & 55.00 & 8.00 & 351275 & 81.00 & 166.00 & 242.00 & 161.00 & 161071 & 2.00 & 2.00 & 4.00 & 2.00 \\
linea $\rightarrow$ arbitrum & & & & &  & 3 & 1330.00 & 1353.00 & 1362.50 & 32.50 & 143954 & 30.00 & 31.00 & 49.00 & 19.00 & 298289 & 70.00 & 167.00 & 262.00 & 192.00 & 153007 & 0.00 & 1.00 & 8.00 & 8.00 \\
linea $\rightarrow$ optimism & & & & &  & & & & &  & 54518 & 47.00 & 48.00 & 52.00 & 5.00 & 169524 & 76.00 & 178.00 & 293.00 & 217.00 & 84983 & 2.00 & 4.00 & 14.00 & 12.00 \\
linea $\rightarrow$ scroll & & & & &  & & & & &  & 32828 & 57.00 & 143.00 & 176.00 & 119.00 & 171559 & 173.00 & 270.00 & 360.00 & 187.00 & 27855 & -1.00 & 0.00 & 1.00 & 2.00 \\
scroll $\rightarrow$ base & & & & &  & & & & &  & 61261 & 62.00 & 90.00 & 93.00 & 31.00 & 144840 & 122.00 & 167.00 & 246.00 & 124.00 & 79877 & 5.00 & 6.00 & 8.00 & 3.00 \\
scroll $\rightarrow$ arbitrum & & & & &  & & & & &  & 57660 & 46.00 & 72.00 & 74.00 & 28.00 & 164736 & 102.00 & 138.00 & 229.00 & 127.00 & 56345 & 3.00 & 3.00 & 4.00 & 1.00 \\
scroll $\rightarrow$ optimism & & & & &  & & & & &  & 22812 & 52.00 & 89.00 & 92.00 & 40.00 & 83714 & 123.00 & 181.00 & 260.00 & 137.00 & 31147 & 5.00 & 5.00 & 8.00 & 3.00 \\
scroll $\rightarrow$ linea & & & & &  & & & & &  & 35445 & 74.00 & 78.00 & 81.00 & 7.00 & 82042 & 119.00 & 200.00 & 270.00 & 151.00 & 47078 & 4.00 & 5.00 & 7.00 & 3.00 \\
\hspace*{0.5cm}\textbf{L2 $\rightarrow$ L2} & 284161 & 1143.00 & 1287.00 & 1468.00 & 325.00 & 1043 & 1140.00 & 1263.00 & 1434.00 & 294.00 & 2489839 & 32.00 & 46.00 & 56.00 & 24.00 & 2991085 & 89.00 & 157.00 & 241.00 & 152.00 & 2760250 & 2.00 & 2.00 & 5.00 & 3.00 \\
\hline
\end{tabular}
\caption{Count, median, and standard deviation of the latency for all cross-chain transactions across multiple cross-chain protocols between Jun 1, 2024 and Dec 31, 2024. Protocols included: CCTP, CCIP, Stargate (Taxi), Stargate (Bus), Across.}
\end{table}

\end{landscape}

%% file: tables/results-cost.tex
\setlength{\tabcolsep}{6pt} 

\begin{landscape}

\begin{table}
\rowcolors{2}{gray!10}{white}
\tiny
\begin{tabular}{lrrrrrrrrrrrrrrrrrrrrrrrrr}
\toprule
 & \multicolumn{5}{c}{\textbf{cctp}} & \multicolumn{5}{c}{\textbf{ccip}} & \multicolumn{5}{c}{\textbf{stargate\_oft}} & \multicolumn{5}{c}{\textbf{stargate\_bus}} & \multicolumn{5}{c}{\textbf{across}} \\
\cmidrule(rl){2-6} \cmidrule(rl){7-11} \cmidrule(rl){12-16} \cmidrule(rl){17-21} \cmidrule(rl){22-26} 
 & n & Q1 & Q2 & Q3 & IQR & n & Q1 & Q2 & Q3 & IQR & n & Q1 & Q2 & Q3 & IQR & n & Q1 & Q2 & Q3 & IQR & n & Q1 & Q2 & Q3 & IQR \\
\hline
ethereum $\rightarrow$ avalanche & 2211 & 2.62 & 5.56 & 11.25 & 8.63 & 113 & 11.24 & 14.87 & 17.51 & 6.27 & 1676 & 3.95 & 7.62 & 12.44 & 8.49 & 374 & 2.29 & 4.47 & 9.42 & 7.13 & & & & &  \\
ethereum $\rightarrow$ bnb & & & & &  & 591 & 10.05 & 12.15 & 16.79 & 6.73 & 9129 & 3.29 & 6.62 & 12.02 & 8.73 & 2945 & 1.72 & 4.08 & 9.68 & 7.96 & & & & &  \\
avalanche $\rightarrow$ ethereum & 4126 & 0.12 & 0.20 & 0.33 & 0.21 & 71 & 18.15 & 23.26 & 30.50 & 12.36 & 1408 & 0.10 & 0.18 & 0.28 & 0.18 & 1085 & 1.62 & 4.50 & 13.05 & 11.43 & & & & &  \\
avalanche $\rightarrow$ bnb & & & & &  & 303 & 2.57 & 2.94 & 3.31 & 0.74 & 6279 & 0.09 & 0.16 & 0.30 & 0.21 & 4219 & 0.29 & 0.43 & 0.71 & 0.43 & & & & &  \\
bnb $\rightarrow$ avalanche & & & & &  & 457 & 2.49 & 2.61 & 2.82 & 0.32 & 6896 & 0.15 & 0.22 & 0.44 & 0.29 & 2726 & 0.34 & 0.47 & 0.82 & 0.48 & & & & &  \\
bnb $\rightarrow$ ethereum & & & & &  & 286 & 18.22 & 25.11 & 33.99 & 15.77 & 13327 & 0.17 & 0.30 & 0.47 & 0.30 & 8684 & 1.67 & 3.58 & 8.61 & 6.94 & & & & &  \\
\hspace*{0.5cm}\textbf{L1 $\rightarrow$ L1} & \textbf{6337} & \textbf{0.16} & \textbf{0.34} & \textbf{2.91} & \textbf{2.75} & \textbf{1821} & \textbf{2.82} & \textbf{9.58} & \textbf{16.82} & \textbf{14.00} & \textbf{38715} & \textbf{0.18} & \textbf{0.40} & \textbf{1.45} & \textbf{1.27} & \textbf{20033} & \textbf{0.57} & \textbf{1.76} & \textbf{5.55} & \textbf{4.99} & & & & &  \\
\hline
ethereum $\rightarrow$ base & 23882 & 1.51 & 3.51 & 7.73 & 6.22 & 461 & 9.32 & 12.87 & 18.71 & 9.39 & 27002 & 2.84 & 5.53 & 9.83 & 6.99 & 16476 & 1.26 & 2.56 & 4.89 & 3.63 & 279910 & 2.04 & 3.47 & 5.80 & 3.76 \\
ethereum $\rightarrow$ arbitrum & 17135 & 2.14 & 4.39 & 8.53 & 6.39 & 443 & 7.94 & 11.40 & 18.42 & 10.48 & 22117 & 1.93 & 4.00 & 7.30 & 5.37 & 20554 & 0.91 & 2.08 & 3.93 & 3.02 & 94264 & 1.41 & 2.98 & 5.68 & 4.26 \\
ethereum $\rightarrow$ optimism & 3964 & 1.85 & 3.88 & 8.58 & 6.73 & 22 & 9.52 & 16.49 & 21.97 & 12.45 & 3656 & 1.59 & 3.43 & 6.36 & 4.77 & 3635 & 0.86 & 1.79 & 3.38 & 2.52 & 23022 & 1.17 & 2.63 & 5.14 & 3.97 \\
ethereum $\rightarrow$ linea & & & & &  & 2 & 8.68 & 8.77 & 8.85 & 0.16 & 2214 & 1.87 & 3.08 & 5.42 & 3.55 & 8017 & 1.01 & 2.06 & 3.17 & 2.16 & 25856 & 1.00 & 1.96 & 3.88 & 2.89 \\
ethereum $\rightarrow$ scroll & & & & &  & & & & &  & 1647 & 2.42 & 4.31 & 7.98 & 5.56 & 6078 & 1.01 & 1.75 & 2.92 & 1.90 & 4467 & 0.68 & 1.89 & 3.64 & 2.97 \\
avalanche $\rightarrow$ base & 14238 & 0.13 & 0.23 & 0.39 & 0.26 & 126 & 1.80 & 2.11 & 2.29 & 0.49 & 8870 & 0.09 & 0.13 & 0.24 & 0.15 & 3349 & 0.16 & 0.26 & 0.41 & 0.25 & & & & &  \\
avalanche $\rightarrow$ arbitrum & 18619 & 0.15 & 0.24 & 0.35 & 0.20 & 110 & 1.87 & 2.20 & 2.38 & 0.51 & 15122 & 0.10 & 0.12 & 0.16 & 0.07 & 9411 & 0.14 & 0.21 & 0.40 & 0.26 & & & & &  \\
avalanche $\rightarrow$ optimism & 5401 & 0.10 & 0.18 & 0.28 & 0.18 & 10 & 0.35 & 0.52 & 0.73 & 0.38 & 4856 & 0.10 & 0.13 & 0.21 & 0.11 & 2406 & 0.15 & 0.21 & 0.36 & 0.22 & & & & &  \\
avalanche $\rightarrow$ linea & & & & &  & 2 & 3.75 & 3.76 & 3.77 & 0.02 & & & & &  & & & & &  & & & & &  \\
avalanche $\rightarrow$ scroll & & & & &  & & & & &  & 536 & 0.27 & 0.35 & 0.50 & 0.22 & 1009 & 0.32 & 0.43 & 0.63 & 0.31 & & & & &  \\
bnb $\rightarrow$ base & & & & &  & 2060 & 0.51 & 0.72 & 0.90 & 0.39 & 38792 & 0.19 & 0.22 & 0.44 & 0.25 & 14569 & 0.24 & 0.34 & 0.61 & 0.37 & & & & &  \\
bnb $\rightarrow$ arbitrum & & & & &  & 358 & 0.53 & 1.28 & 2.05 & 1.52 & 43525 & 0.12 & 0.20 & 0.29 & 0.17 & 31172 & 0.15 & 0.24 & 0.36 & 0.21 & & & & &  \\
bnb $\rightarrow$ optimism & & & & &  & 2 & 0.36 & 0.37 & 0.39 & 0.03 & 13400 & 0.12 & 0.20 & 0.28 & 0.16 & 6966 & 0.15 & 0.22 & 0.33 & 0.18 & & & & &  \\
bnb $\rightarrow$ linea & & & & &  & 2 & 4.37 & 4.37 & 4.37 & 0.00 & & & & &  & & & & &  & & & & &  \\
bnb $\rightarrow$ scroll & & & & &  & & & & &  & 955 & 0.32 & 0.41 & 0.75 & 0.43 & 2305 & 0.35 & 0.47 & 0.69 & 0.34 & & & & &  \\
\hspace*{0.5cm}\textbf{L1 $\rightarrow$ L2} & \textbf{83239} & \textbf{0.24} & \textbf{0.82} & \textbf{4.31} & \textbf{4.07} & \textbf{3598} & \textbf{0.59} & \textbf{0.93} & \textbf{4.51} & \textbf{3.92} & \textbf{182692} & \textbf{0.15} & \textbf{0.28} & \textbf{1.83} & \textbf{1.68} & \textbf{125947} & \textbf{0.24} & \textbf{0.53} & \textbf{1.96} & \textbf{1.72} & \textbf{427519} & \textbf{1.73} & \textbf{3.22} & \textbf{5.62} & \textbf{3.89} \\
\hline
arbitrum $\rightarrow$ avalanche & 9209 & 0.01 & 0.01 & 0.02 & 0.01 & 305 & 2.32 & 2.39 & 2.46 & 0.14 & 50579 & 0.01 & 0.01 & 0.02 & 0.02 & 5834 & 0.18 & 0.31 & 0.84 & 0.66 & & & & &  \\
arbitrum $\rightarrow$ ethereum & 13679 & 0.01 & 0.01 & 0.02 & 0.01 & 363 & 16.69 & 23.79 & 32.89 & 16.20 & 46740 & 0.01 & 0.01 & 0.02 & 0.01 & 69393 & 0.69 & 1.62 & 3.46 & 2.77 & 84100 & 2.70 & 5.57 & 9.83 & 7.13 \\
arbitrum $\rightarrow$ bnb & & & & &  & 1041 & 1.32 & 2.21 & 2.91 & 1.59 & 155311 & 0.01 & 0.01 & 0.64 & 0.63 & 21385 & 0.13 & 0.17 & 0.32 & 0.19 & & & & &  \\
optimism $\rightarrow$ avalanche & 3819 & 0.00 & 0.00 & 0.03 & 0.03 & 3 & -- & -- & -- & -- & 5487 & 0.00 & 0.00 & 0.00 & 0.00 & 2173 & 0.18 & 0.28 & 0.86 & 0.68 & & & & &  \\
optimism $\rightarrow$ ethereum & 3267 & 0.00 & 0.00 & 0.01 & 0.01 & 18 & -- & -- & -- & -- & 4406 & 0.00 & 0.00 & 0.00 & 0.00 & 25077 & 0.57 & 1.20 & 2.56 & 1.99 & 23771 & 2.29 & 4.59 & 8.12 & 5.83 \\
optimism $\rightarrow$ bnb & & & & &  & 1 & -- & -- & -- & -- & 11765 & 0.00 & 0.00 & 0.00 & 0.00 & 8578 & 0.12 & 0.15 & 0.22 & 0.10 & & & & &  \\
base $\rightarrow$ avalanche & 8598 & 0.00 & 0.01 & 0.02 & 0.01 & 308 & 2.32 & 2.38 & 2.45 & 0.13 & 8046 & 0.00 & 0.00 & 0.01 & 0.01 & 2326 & 0.16 & 0.28 & 0.87 & 0.71 & & & & &  \\
base $\rightarrow$ ethereum & 22272 & 0.00 & 0.00 & 0.01 & 0.01 & 353 & 19.42 & 26.64 & 36.63 & 17.21 & 14987 & 0.00 & 0.00 & 0.01 & 0.01 & 56950 & 0.62 & 1.54 & 3.76 & 3.14 & 140641 & 3.95 & 6.71 & 10.72 & 6.77 \\
base $\rightarrow$ bnb & & & & &  & 676 & 1.36 & 2.09 & 2.85 & 1.48 & 26418 & 0.00 & 0.01 & 0.02 & 0.01 & 10240 & 0.13 & 0.17 & 0.30 & 0.16 & & & & &  \\
linea $\rightarrow$ avalanche & & & & &  & 2 & 0.10 & 0.10 & 0.10 & 0.01 & & & & &  & & & & &  & & & & &  \\
linea $\rightarrow$ ethereum & & & & &  & 2 & 0.09 & 0.10 & 0.10 & 0.01 & 4279 & 0.01 & 0.02 & 0.05 & 0.03 & 67065 & 1.41 & 2.50 & 4.44 & 3.03 & 17289 & 2.14 & 3.95 & 7.19 & 5.05 \\
linea $\rightarrow$ bnb & & & & &  & 2 & 0.09 & 0.10 & 0.10 & 0.01 & & & & &  & & & & &  & & & & &  \\
scroll $\rightarrow$ avalanche & & & & &  & & & & &  & 649 & 0.06 & 0.09 & 0.15 & 0.10 & 855 & 0.18 & 0.24 & 0.38 & 0.20 & & & & &  \\
scroll $\rightarrow$ ethereum & & & & &  & & & & &  & 2894 & 0.05 & 0.07 & 0.11 & 0.06 & 27824 & 1.18 & 2.16 & 3.53 & 2.35 & 7679 & 2.62 & 4.69 & 6.88 & 4.26 \\
scroll $\rightarrow$ bnb & & & & &  & & & & &  & 1961 & 0.06 & 0.09 & 0.13 & 0.07 & 3745 & 0.16 & 0.20 & 0.28 & 0.12 & & & & &  \\
\hspace*{0.5cm}\textbf{L2 $\rightarrow$ L1} & \textbf{60844} & \textbf{0.00} & \textbf{0.01} & \textbf{0.01} & \textbf{0.01} & \textbf{3074} & \textbf{2.02} & \textbf{2.52} & \textbf{3.37} & \textbf{1.35} & \textbf{333522} & \textbf{0.01} & \textbf{0.01} & \textbf{0.04} & \textbf{0.04} & \textbf{301445} & \textbf{0.50} & \textbf{1.45} & \textbf{3.11} & \textbf{2.61} & \textbf{273480} & \textbf{3.04} & \textbf{5.93} & \textbf{9.98} & \textbf{6.94} \\
\hline
arbitrum $\rightarrow$ base & 76498 & 0.02 & 0.03 & 0.04 & 0.02 & 450 & 0.34 & 0.44 & 1.00 & 0.66 & 576116 & 0.01 & 0.01 & 0.01 & 0.01 & 208847 & 0.02 & 0.03 & 0.06 & 0.03 & 439916 & 0.01 & 0.02 & 0.05 & 0.04 \\
arbitrum $\rightarrow$ optimism & 24770 & 0.01 & 0.02 & 0.03 & 0.02 & 28 & 0.27 & 0.31 & 0.35 & 0.09 & 411998 & 0.01 & 0.01 & 0.02 & 0.01 & 133141 & 0.02 & 0.03 & 0.04 & 0.02 & 141592 & 0.01 & 0.03 & 0.09 & 0.07 \\
arbitrum $\rightarrow$ linea & & & & &  & 4 & 3.53 & 3.61 & 3.69 & 0.16 & 36935 & 0.01 & 0.01 & 0.01 & 0.01 & 69355 & 0.08 & 0.11 & 0.16 & 0.09 & 197122 & 0.07 & 0.09 & 0.14 & 0.07 \\
arbitrum $\rightarrow$ scroll & & & & &  & & & & &  & 24070 & 0.01 & 0.01 & 0.02 & 0.01 & 110142 & 0.17 & 0.19 & 0.28 & 0.11 & 43424 & 0.07 & 0.11 & 0.17 & 0.09 \\
optimism $\rightarrow$ base & 38762 & 0.00 & 0.00 & 0.02 & 0.02 & 68 & -- & -- & -- & -- & 181804 & 0.00 & 0.00 & 0.00 & 0.00 & 175736 & 0.02 & 0.03 & 0.04 & 0.02 & 264441 & 0.01 & 0.02 & 0.05 & 0.04 \\
optimism $\rightarrow$ arbitrum & 35364 & 0.00 & 0.00 & 0.05 & 0.05 & 23 & -- & -- & -- & -- & 170302 & 0.00 & 0.00 & 0.00 & 0.00 & 167702 & 0.02 & 0.03 & 0.04 & 0.02 & 161243 & 0.02 & 0.04 & 0.09 & 0.06 \\
optimism $\rightarrow$ linea & & & & &  & & & & &  & 22221 & 0.00 & 0.00 & 0.00 & 0.00 & 53572 & 0.09 & 0.13 & 0.18 & 0.08 & 147027 & 0.07 & 0.09 & 0.14 & 0.07 \\
optimism $\rightarrow$ scroll & & & & &  & & & & &  & 11776 & 0.00 & 0.00 & 0.05 & 0.04 & 77823 & 0.17 & 0.18 & 0.26 & 0.09 & 25391 & 0.07 & 0.11 & 0.16 & 0.09 \\
base $\rightarrow$ arbitrum & 71532 & 0.01 & 0.01 & 0.02 & 0.02 & 379 & 0.35 & 0.45 & 1.79 & 1.44 & 250438 & 0.00 & 0.00 & 0.01 & 0.01 & 198497 & 0.03 & 0.03 & 0.06 & 0.03 & 278554 & 0.02 & 0.04 & 0.08 & 0.06 \\
base $\rightarrow$ optimism & 37235 & 0.00 & 0.01 & 0.02 & 0.01 & 84 & 0.27 & 0.29 & 0.35 & 0.07 & 122690 & 0.00 & 0.01 & 0.01 & 0.01 & 139906 & 0.02 & 0.02 & 0.03 & 0.02 & 199692 & 0.01 & 0.02 & 0.07 & 0.06 \\
base $\rightarrow$ linea & & & & &  & 2 & 3.66 & 3.67 & 3.67 & 0.01 & 55973 & 0.00 & 0.00 & 0.01 & 0.00 & 87467 & 0.08 & 0.11 & 0.17 & 0.09 & 170460 & 0.06 & 0.08 & 0.13 & 0.06 \\
base $\rightarrow$ scroll & & & & &  & & & & &  & 22668 & 0.00 & 0.01 & 0.01 & 0.01 & 102918 & 0.18 & 0.20 & 0.31 & 0.13 & 50025 & 0.07 & 0.10 & 0.15 & 0.09 \\
linea $\rightarrow$ base & & & & &  & 2 & 0.09 & 0.09 & 0.09 & 0.00 & 194370 & 0.01 & 0.02 & 0.04 & 0.03 & 351275 & 0.03 & 0.05 & 0.08 & 0.04 & 161071 & 0.03 & 0.05 & 0.12 & 0.09 \\
linea $\rightarrow$ arbitrum & & & & &  & 3 & 0.09 & 0.09 & 0.10 & 0.01 & 143954 & 0.01 & 0.02 & 0.04 & 0.03 & 298289 & 0.04 & 0.05 & 0.09 & 0.05 & 153007 & 0.04 & 0.07 & 0.14 & 0.10 \\
linea $\rightarrow$ optimism & & & & &  & & & & &  & 54518 & 0.01 & 0.02 & 0.05 & 0.03 & 169524 & 0.03 & 0.05 & 0.06 & 0.03 & 84983 & 0.03 & 0.08 & 0.12 & 0.09 \\
linea $\rightarrow$ scroll & & & & &  & & & & &  & 32828 & 0.01 & 0.02 & 0.05 & 0.03 & 171559 & 0.20 & 0.23 & 0.31 & 0.11 & 27855 & 0.09 & 0.12 & 0.19 & 0.10 \\
scroll $\rightarrow$ base & & & & &  & & & & &  & 61261 & 0.04 & 0.05 & 0.08 & 0.05 & 144840 & 0.04 & 0.06 & 0.09 & 0.05 & 79877 & 0.03 & 0.04 & 0.08 & 0.05 \\
scroll $\rightarrow$ arbitrum & & & & &  & & & & &  & 57660 & 0.04 & 0.06 & 0.09 & 0.06 & 164736 & 0.05 & 0.06 & 0.09 & 0.04 & 56345 & 0.04 & 0.07 & 0.12 & 0.08 \\
scroll $\rightarrow$ optimism & & & & &  & & & & &  & 22812 & 0.04 & 0.06 & 0.09 & 0.05 & 83714 & 0.04 & 0.04 & 0.06 & 0.02 & 31147 & 0.02 & 0.04 & 0.09 & 0.07 \\
scroll $\rightarrow$ linea & & & & &  & & & & &  & 35445 & 0.02 & 0.03 & 0.05 & 0.03 & 82042 & 0.11 & 0.15 & 0.21 & 0.10 & 47078 & 0.08 & 0.12 & 0.18 & 0.10 \\
\hspace*{0.5cm}\textbf{L2 $\rightarrow$ L2} & \textbf{284161} & \textbf{0.00} & \textbf{0.01} & \textbf{0.03} & \textbf{0.03} & \textbf{1043} & \textbf{0.33} & \textbf{0.42} & \textbf{0.86} & \textbf{0.53} & \textbf{2489839} & \textbf{0.00} & \textbf{0.01} & \textbf{0.02} & \textbf{0.02} & \textbf{2991085} & \textbf{0.03} & \textbf{0.05} & \textbf{0.14} & \textbf{0.11} & \textbf{2760250} & \textbf{0.02} & \textbf{0.05} & \textbf{0.11} & \textbf{0.08} \\
\hline
\end{tabular}
\caption{Count, median, and standard deviation of the cost for all cross-chain transactions across multiple cross-chain protocols between X 2024 and 1 Jan 2025. Protocols included: CCTP, CCIP, Stargate (OFT), Stargate (Bus), Across.}
\end{table}

\end{landscape}